\documentclass{aa}
\usepackage{lscape}
\usepackage{longtable}
\usepackage{txfonts}
\usepackage{graphicx}
\usepackage{url}

\usepackage{soul} 
\usepackage{color}

\def\sec{\ifmmode {}^{\prime\prime}\else ${}^{\prime\prime}$\fi~}

\def\magdot{\ifmmode {}^{\rm m}\!\!\!.\, \else ${}^{\rm m}\!\!\!.\,$\fi}
\def\daydot{\ifmmode {}^{\rm d}\!\!\!.\, \else ${}^{\rm d}\!\!\!.\,$\fi}
\def\asec{\ifmmode ^{\prime\prime}\else$^{\prime\prime}$\fi}

\usepackage{color}

\newcommand{\dfrac}[2]{\frac{\displaystyle #1}{\displaystyle #2}}

\begin{document}

\title{The Nainital-Cape Survey-IV }
\author{Santosh Joshi\inst{1}, P. Martinez\inst{2,3}, Sowgata Chowdhury\inst{1,4}, N. K. Chakradhari\inst{5}, Y. C. Joshi\inst{1}, P. van Heerden\inst{3}, \\
T. Medupe\inst{6}, Yerra Bharat Kumar\inst{7}, R. B. Kuhn\inst{3}
}
\offprints{Santosh Joshi (santosh@aries.res.in)}

\authorrunning{S. Joshi et al.}

\institute{
Aryabhatta Research Institute of Observational Sciences, Manora peak,
Nainital-263129, India \\
\email{santosh@aries.res.in}
\and
SpaceLab, Department of Electrical Engineering, University of Cape Town, Private Bag X3, Rondebosch 7701, South Africa 
\and
South African Astronomical Observatory, P.O. Box 9, Observatory 7935, South Africa
\and
Department of Physics, Christ University, Hosur Road, Bangalore - 560029, Karnataka, India
\and
School of Studies in Physics and Astrophysics, Pt Ravishankar Shukla University, Raipur 492 010, India
\and
Department of Physics, University of the North-West, Private Bag X2046, Mmabatho 2735, South Africa
\and
National Astronomical Observatories, Chinese Academy of Sciences, 20A Datun Road, Chaoyang District, Beijing, China
}
\date{Received; accepted}

\authorrunning{S. Joshi et al.}
\titlerunning{The Nainital-Cape Survey-IV}

\abstract
{The Nainital-Cape survey is a dedicated ongoing survey programme to search for and study  pulsational variability in chemically peculiar (CP) stars to understand their internal structure and evolution. }
{The main aims of this survey are to find new pulsating Ap and Am stars in the northern and southern hemisphere and to perform asteroseismic studies of these new pulsators.}
{The survey is conducted using high-speed photometry.  The candidate stars were selected on the basis of having Str\"omgren photometric indices similar to those of known pulsating CP stars.}
{Over the last decade a total of 337 candidate pulsating CP stars were observed for the Nainital-Cape survey, making it one of the longest ground-based surveys for pulsation in CP stars in terms of time span and sample size. The previous papers of this series presented seven new pulsating variables and 229 null results. In this paper we present the light curves, frequency spectra and the various astrophysical parameters of the 108 additional CP stars observed since the last reported results. We have also tabulated the basic physical parameters of the known roAp stars. As part of establishing the detection limits in the Nainital-Cape survey, we investigated the scintillation noise level at the two observing sites used in this survey, Sutherland and Nainital, by comparing the combined frequency spectra stars observed from each location. Our analysis shows that both the sites permit the detection of variations on the order of 0.6 milli-magnitude (mmag) in the frequency range 1-4 mHz, with Sutherland being marginally better, on average.}
{}


\maketitle

\section{Introduction}
\label{sec:intro}

A chemically peculiar (CP) star can be  distinguished from a chemically normal star by its spectrum, where anomalies can be seen on a visual inspection of low-dispersion
spectra. The optical spectra of the CP stars exhibit normal hydrogen lines combined with 
enhanced silicon, metal, and/or rare-earth lines and weak calcium lines. The chemical peculiarities in these stars result from the diffusion process (Michaud \cite{micha70}; Michaud et al. \cite{micha81}; Babel \cite{babe92}; Richer et al. \cite{rich00}). Chemical elements with many lines near flux maximum, such as iron peak and rare earth elements, are brought up to the surface by the dominance of radiation pressure over gravity in the radiative envelopes of these stars, causing an apparent over-abundance of such elements. 
The elements with few lines near the flux maximum settle gravitationally and 
appear to be under-abundant. Slow rotation is thus a basic condition to operate the diffusion process in CP stars. The CP stars are found on the main-sequence between spectral types B2 and F5, from the zero-age main-sequence (ZAMS) to the
terminal-age main-sequence (TAMS), and have masses ranging from 1.5 to
about 7 $M_{\sun}$. 

Based on their spectroscopic characteristics, Preston (\cite{pres74}) 
divided the CP stars into  following groups : Am/Fm (CP1), Ap/Bp (CP2), 
Hg-Mn (CP3), He weak and He strong (CP4) stars. Renson \& Manfroid  
(\cite{rens09}) compiled  an up-to-date catalogue of 8205 CP stars. A 
subset of Ap and Am stars shows photometric variability with periods ranging 
from a few minutes to a few hours, and are the focus of the Nainital - Cape survey. 
   
The Am/Fm stars are relatively cool stars of spectral type F5-A8,  with 
temperatures ranging from 6500\,K to 10000\,K. The spectra of these stars exhibit 
an under-abundance (weak lines) of
Ca or Sc (or of both elements) and over-abundance (strong lines) of Sr, Eu and 
other rare-earth elements. Some of the members of this group 
show  $\delta$ Sct-type pulsational variability (Joshi et al. \cite{josh03}, 
\cite{josh06}, \cite{josh09}; Smalley et al. \cite{smal11}; Catanzaro \& Ripepi \cite{cata14}; Hou et al. \cite{hou15}). The Am stars 
rotate slower than chemically normal A-type stars and 
the frequency of binarity among these stars  is much higher than among normal 
stars of the same mass (Abt \& Golson \cite{abt62}, Abt \& Snowden 
\cite{abt73}). It is well understood that these stars do not exhibit strong global magnetic fields, however based on the observations from Kepler space mission, Balona et al. (\cite{balo15}) found flares in two Am stars, which strongly suggests that at least some Am stars possess the significant magnetic fields. 
 
The Ap/Bp stars have effective temperatures in the range of 6400\,K to 
15000\,K.  These stars exhibit the most conspicuous chemical anomalies of all the CP 
stars: enhanced lines of
some elements, particularly Si, Cr, Sr, Mn, Fe, Eu, Gd, and Ce (overabundant by 
up to a factor of $10^6$), and weak lines of light elements (under-abundant by 
a factor of $10^{-2}$). The Ap stars show
low rotation velocities with $v_e \sin i$ usually not exceeding 100 kms$^{-1}$. These 
stars have strong global magnetic fields with an intensity ranging from 
hundreds of gauss to tens of kilogauss. 

 The coolest  sub-group of Ap stars (6400\,K $\le T_{eff} \le$ 8700\,K) located 
near the main-sequence (MS) part of the classical instability strip, are known as 
roAp stars. 
Since the  discovery of first roAp star HD\,101065 (Kurtz \cite{kurt78}), 61 
other members of this class have been discovered (Smalley et al. \cite{smal15}). 
The roAp stars show pulsational variability in both the broad photometric bands 
and in narrow spectral lines. These pulsations are characterized as 
high-overtone, low-degree $p$-modes with typical periods between 5.6 min and 23.6 min and photometric amplitudes ranging from a few micro-magnitudes ($\mu$mag) up to tens of milli-magnitudes (mmag) and radial velocity (RV) amplitudes ranging from a few ms$^{-1}$ to kms$^{-1}$. The roAp stars  possess strong magnetic fields with typical strengths of a few kG to tens of kG (Hubrig et al. \cite{hubr12}) with over-abundances of some rare earth 
elements that can exceed the solar value by $10^{6}$  (Ryabchikova et al. 
\cite{ryab04}). To date, there have been no roAp stars found in close binary
systems though a few Ap stars are in close binaries. The roAp stars are among the more challenging MS stars to model due to their  pulsations in the combined presence of a strong global magnetic field together with element segregation and stratification, but at the same time they can be considered as a stellar atomic physics laboratory.

The pulsation frequency spectrum of some of the roAp stars shows frequency 
multiplets with spacings corresponding to the frequency of rotation of the star. This 
phenomenon can be  explained using the oblique pulsator model (Kurtz \cite{kurt82}), in which the pulsation axis is aligned with the axis of the magnetic field, which
is assumed to be roughly a dipole inclined with respect to the axis of
rotation. As a star rotates, the observed aspect of the pulsation changes,
leading to amplitude modulation and, in some cases, phase modulation.
The driving mechanism of the pulsations in roAp stars is thought to be the classical $\kappa$-mechanism operating in the partial hydrogen ionization zone (Balmforth et al. \cite{balm01}). Cunha \& Gough (\cite{cunh01}) suggested an alternative excitation mechanism for roAp stars where pulsation is driven by the turbulent pressure in the convection zone.   

Some roAp stars have highly stable pulsation frequencies and amplitudes, even on time scales of years while other roAp stars show frequency and amplitude variations on time scales as short as hours (Medupe et al. \cite{medu15}). Whether this is a result of driving and damping, mode coupling or some instability is not known. It is important to know where in the roAp instability strip the stable and unstable pulsators lie.

The Kepler mission, launched in 2009 with the aim to detect and
characterize Earth-sized planets in the habitable zone, has revolutionized 
our ability to detect and study very low-amplitude light variations of the order of a few $\mu$-mag  in rather faint stars (Koch et al. \cite{koch10}). The Kepler mission has enabled the discovery of five roAp stars, all which have pulsation amplitudes
much below the detection limits of ground-based photometry.

While initially roAp stars were discovered and studied with photometric methods, time-resolved spectroscopy has allowed the study of wider physical aspects of the pulsating stellar atmosphere. The rapid radial velocity variations of spectral lines of certain chemical elements allow us to sample the velocity field in the stellar atmosphere as a function of atmospheric depth. Of the 61 known roAp stars, about a quarter of them were discovered using spectroscopic methods. A combination of simultaneous spectroscopy and photometry constitutes the most sophisticated asteroseismic data set for any roAp star. The observed phase lag between the variations in luminosity and in RV is an important parameter
for modeling the stellar structure. 

Similar to other pulsating stars, the roAp stars are also excellent asteroseismic candidates through which one can compare the observed frequency spectrum  to the asymptotic pulsation theory and then obtain information about the spherical harmonic degrees of the pulsation modes, the distortion of the modes from normal modes, atmospheric structures, evolutionary status and the geometry of the magnetic field. Using such information one can derive the various physical parameters such as rotation periods, temperatures, luminosities, radii and their masses (see Joshi \& Joshi (\cite{josh15}) for a recent review on asteroseismology of pulsating stars). Although the extent of the roAp phenomenon has been fairly well delineated in photometric and spectroscopic terms, there is as yet no known combination of these (and other) observable parameters that can be used as a predictor of pulsation in any given Ap star. In other words, one can have two Ap stars that are apparently similar in all observable parameters, where one is a pulsating roAp star and the other has no detectable pulsations and is a so-called ``noAp'' star.  

The Nainital-Cape survey was initiated in 1999 by the Aryabhatta Research Institute of Observational Sciences (ARIES) at Manora Peak, Nainital, India,
 and the South African Astronomical Observatory (SAAO) in Sutherland to search for pulsations in CP stars. The goals of the survey were: (i) to increase the number of known pulsating CP stars; (ii) to determine the observational limits of the roAp phenomenon; and (iii) to broaden the number and distribution (in parameter space) of established constant (noAp) stars, so as to shed some light on what distinguishes the pulsating from the apparently constant CP stars of similar spectral type and other observable physical parameters. This is the only survey of its kind which was conducted from both the northern and southern hemisphere. 
The first three papers of this  survey described the 
scope and methods of the survey and reported the discovery of pulsations in 
several CP stars (Martinez et al. \cite{mart01}; Paper-I, Joshi et al. 
\cite{josh06}; Paper-II, Joshi et al. \cite{josh09}: Paper-III). The present 
paper is the fourth in this series and presents the null results  obtained for 
108 stars observed during the period of 2006 to 2009.  

Similar to other papers of this series, the present paper is also based on photoelectric photometry of the sample stars and is organized as follows: the target selection, observations and data reduction procedures are described in Section \ref{obs}, followed by the frequency analysis of the time series photometric data in Section \ref{frqanalysis}.  In Sec. \ref{noise}, the observational limits for the detection of light variations at the ARIES and SAAO sites are discussed. The stars classified as null results and their basic astrophysical parameters are given in Sec. \ref{nulls}. In Sec. \ref{comp}, we provide the basic physical parameters of all the currently known roAp stars. In this section, we also compare the evolutionary status of the known roAp stars to the sample of stars observed under the Nainital-Cape survey. The statistics of several surveys to search for  new roAp stars are discussed in Sec. \ref{survey}. Finally, we outline the conclusions drawn from our study in Sec. \ref{conc}.   

\section{Target Selection, Observations and Data Reduction}
\label{obs}

\subsection{Selection Criteria}
Following the target selection strategy of Martinez et al. (\cite{mart91}), the primary source 
of candidates for the Nainital-Cape survey was the subset of CP stars with Str\"omgren photometric indices 
similar to those of the known roAp stars. In this range, we have also found many Am stars and included them in the list of targets. Apart from the sources of target mentioned in Martinez et al. (\cite{mart91}), we also included Ap/Am stars from Renson et al. (\cite{rens91}) and magnetic stars from Bychkov et al. (\cite{Bychkov03}).

On the basis of the Str\"omgren photometric indices of known roAp stars (see Table \ref{roap01}), we have revised the range of indices that encompass the roAp phenomenon:  

\begin{center}
0.082 $\leq b-y \leq$ 0.431 \\
0.178 $\leq m_1 \leq$ 0.387 \\
-0.204$\leq \delta m_1 \leq$ 0.012\\
0.002 $\leq c_1 \leq$ 0.870 \\
-0.370$\leq  \delta c_1 \leq$ 0.031\\
2.64 $\leq \beta \leq$ 2.88  \\
\end{center}

\noindent where $b-y$ is the color index and $\beta$ measures the 
strength of the $H_\beta$ line, which  is indicator of temperature for stars in  the
spectral range from around A3 to F2. The $m_1$ and $c_1$ indices indicate enhanced 
metallicity and increased line blanketing, respectively. The parameters $\delta 
m_1$ and $\delta c_1$ measure the blanketing difference and Balmer 
discontinuity relative to the ZAMS for a given $\beta$, respectively. Indices in the 
ranges given above are not an unambiguous indicator of roAp pulsation, although 
they serve to narrow down the field of candidates to the most promising subset. 
It is interesting to note that, whereas previously the roAp phenomenon seemed 
to be confined to the temperature range of the $\delta$ Scuti instability strip, 
it now appears that the roAp instability strip has a considerably cooler red 
edge, well into the F-type stars (see Fig. \ref{HR}).  As can be seen by the 
paucity of cooler stars tested for pulsation, this is an area for future work, 
to establish more firmly the cool edge of the roAp instability strip.

\subsection{Photometric Observations}

For many roAp stars, the pulsational photometric variations have amplitudes less than twenty mmag. The detection of such low-amplitude variations demands high-precision photometric observations that can be attained with fast photometers mounted on small telescopes at  observing sites such as ARIES Nainital in India and SAAO Sutherland South Africa. The ARIES observations presented in this paper were acquired using the ARIES high-speed photoelectric photometer (Ashoka et al. \cite{ashoka01}) attached to the 1.04-m Sampurnanand telescope at ARIES. The SAAO observations were acquired using the Modular Photometer attached to the 0.5-m telescope and the University of Cape Town photometer attached to the 0.75-m and 1.0-m telescopes at the Sutherland site of SAAO.

Each star was observed in high-speed photometric mode with continuous 10-sec 
 integrations through a  Johnson $B$ filter. The observations were acquired in a 
single-channel mode (i.e. no simultaneous comparison star observations), with 
occasional interruptions to measure the sky background, depending on the phase 
and position of the moon. To minimize the  effects  of seeing  fluctuations  and 
tracking  errors, we selected a photometric aperture  of 30$^{\prime\prime}$. 
Each target was observed continuously for  1-3 hours at a time. Since the 
amplitudes of the rapid photometric oscillations in roAp stars exhibit 
modulation  due to rotation and also due to interference among frequencies of different pulsation modes, a null detection for pulsation may be obtained simply due to a coincidence of the 
timing of the observations. Hence, each candidate was observed several times. 

\subsection{Data Reduction}

The data reduction process began with a  visual inspection of the light curve 
to identify and remove  obviously bad data points, followed by  correction for 
coincidence counting losses, subtraction of
the interpolated sky background, and correction for the mean atmospheric 
extinction. After applying these corrections, the time of the mid-point of the each 
observation was converted into a heliocentric Julian date (HJD) with an accuracy of 
$10^{-5}$ day ($\sim$1 sec). The reduced data comprise a time-series of HJD and $\Delta B$ 
magnitude with respect to the mean of the light curve. 

\section{Frequency Analysis}
\label{frqanalysis}

 A fast algorithm (Kurtz \cite{kurt85}) based on Deeming's discrete Fourier transform (DFT) for unequally spaced data (Deeming  \cite{deem75}) was used to calculate the Fourier Transformation. The light curves were also inspected visually for evidence of $\delta$\,Sct oscillations with periods of a few tens of minutes and longer. On these time-scales, single-channel photometric data are affected by sky transparency variations and it is not always possible to distinguish  between oscillations in the star and variations in sky transparency. This is where the comparison of data of the same star acquired under different conditions on different nights is helpful for confirming the tentative detection of coherent oscillations in a given light curve. 

After visual inspection of the light curves to search for indications of $\delta$\,Sct pulsations in a given light curve on time-scales longer than about half an hour, we removed the sky transparency variations from the DFT data in order to reduce the overall noise level to approximately the scintillation noise. This 
is practicable for single-channel data because, on good photometric nights, the roAp oscillation frequencies are generally well resolved from the sky transparency variations. To remove the effect of sky transparency variations, the DFT data were prewhitened to remove signals with frequencies in the range 0 -- 0.9\,mHz, which is the frequency range commonly affected by sky transparency variations in single-channel photometric data. These frequencies were removed one at a time until the noise level in the DFT of the residuals approximated a white noise spectrum. Depending on the stability of the photometric transparency of a given night, it was generally possible to correct for the effects of sky transparency by removing  3-5 frequencies  in the above-mentioned frequency range. 

The first and second panels of Fig. \ref{null1} show the  light curves of the candidate stars filtered for low-frequency sky transparency variations.  The third and fourth panels show the pre-whitened amplitude spectra of the sample stars filtered for low-frequency sky transparency variations.

\section{Noise Level Characterization}
\label{noise}

The detection limit for photometric variability depends upon the atmospheric noise, which consists of scintillation noise and sky-transparency variations, and the photon noise. For the brighter ($\sim$ 10 mag) stars, the atmospheric scintillation noise dominates over the photon noise and is one of the fundamental factors limiting the precision of ground-based photometry. In order to characterize the two observing sites used in the Nainital-Cape survey and to put constraints on the detection limits for low-amplitude variability, we estimated the observational as well as the theoretical scintillation noise values for both the sites.

Given a telescope's altitude, mirror diameter, and the observational exposure time and airmass, one can find the contribution of scintillation noise in photometric measurements using Young's approximation (Young \cite{youn67,youn74}). Using this scaling relation, it is possible to compare the level of scintillation noise at different observatory sites. Although the precise amount of scintillation changes from night to night, Young's scaling relation appears to hold quite well for telescope apertures up to 4-m, and for quite different sites (Kjeldsen \& Frandsen \cite{kjel91}; Gilliland \& Brown \cite{gill92}; Gilliland et al. \cite{gill93}). However recent studies by Kornilov et al. (\cite{korni12}) and Osborn et. al (\cite{osb15}) showed that this equation tends to underestimate the median scintillation noise at several major observatories around the world. Osborn et. al. (\cite{osb15}) presented a modified form of Young's approximation (equation \ref{scieq}) that uses empirical correction coefficients to give more reliable estimates of the scintillation noise at a range of astronomical sites:

\begin{equation}
\sigma_{Y}^2 = 10\times10^{-6}C_Y^2 D^{-4/3}t^{-1}(\cos{\gamma})^{-3} \exp{\left(-2h_\mathrm{obs}/H\right)},
\label{scieq}
\end{equation}
where $C_Y$ is the empirical coefficient, $D$ is the diameter of the telescope, $t$ is the exposure time of the observation, $\gamma$ is the zenith distance, $h_\mathrm{obs}$ is the altitude of the observatory and $H$ the scale height of the atmospheric turbulence, which is generally accepted to be approximately 8000-m. All parameters are in standard SI units. The empirical coefficients $C_Y$ for the major observatories around the world are listed by Osborne et al. (\cite{osb15}).

\begin{figure}[h]
\includegraphics[width=0.95\linewidth]{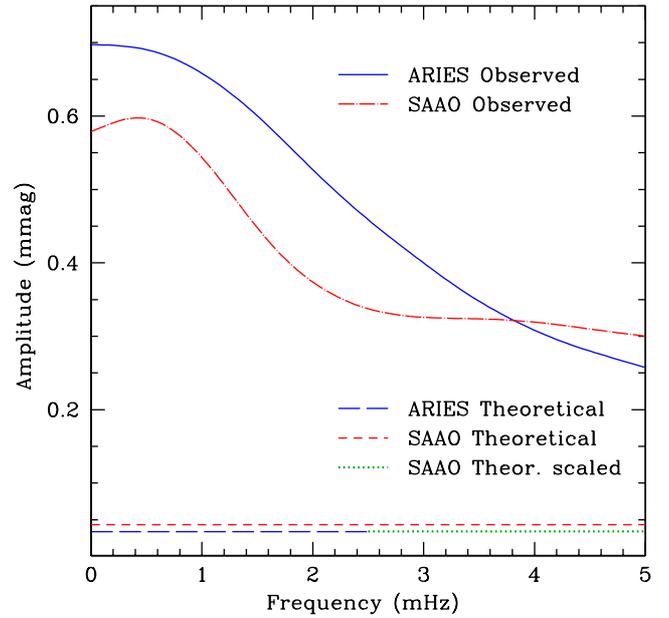}
\caption{{The noise characteristics for the ARIES site at Nainital and the SAAO Sutherland site. The acspline-fitted curve of ARIES and SAAO amplitude spectra are shown in solid blue and dot-dashed red curve, respectively. The theoretical scintillation noise levels of ARIES and SAAO are shown by blue long-dashed and red small-dashed  horizontal lines, respectively, and the scintillation noise level of SAAO (scaled to 104cm diameter) is also shown in green dotted horizontal line.}}

\label{noise_plot}
\end{figure}

The theoretical values of scintillation noise for Sutherland and Nainital were estimated using equation \ref{scieq}. The scintillation noise in terms of amplitude was obtained by taking the square root of ${\sigma_{Y}}$. However, in order to compare the two sites having different telescope diameters, we have to scale the theoretical values. Therefore, the theoretical scintillation noise for SAAO (50-cm telescope) was scaled to the aperture of the ARIES telescope (104-cm) using the same relation. The input parameters used to estimate the theoretical scintillation noise are : height (ARIES: 1958-m, SAAO : 1798-m), sec(Z) (airmass) : 1, $C_Y$ : $1.5$, integration time : 10-sec. The estimated scintillation values of ARIES (D:104-cm) and SAAO (D:50-cm) are 0.0338 mmag and 0.0433 mmag, respectively. The scaled value of the scintillation noise for SAAO (scaled to 104-cm) is 0.0340 mmag. Fig. \ref{noise_plot} shows the theoretical noise levels for the ARIES and SAAO sites (both scaled and unscaled).

Since the observations in the Nainital-Cape survey were carried out over many nights, and in a variety of atmospheric conditions, the noise levels in the Fourier spectra of the individual light curves are expected to be higher than the theoretical scintillation values for each site, and they are also not expected to be white noise. In order to estimate the observational values of the noise in our amplitude spectra as a function of frequency, we first transformed the time-series data of stars observed from ARIES during 2006-2009 and from SAAO during 2006-2007 into their individual periodograms. We then combined all the periodograms from each site into a single pseudo-periodogram and fitted an acspline function to obtain the average estimated noise profile as a function of frequency. These observational noise curves are shown in Fig. \ref{noise_plot} in solid blue for ARIES and dot-dashed red for SAAO. These noise profiles provide a useful first check of the significance of possible oscillation frequencies identified in the Fourier spectra in Fig. 3 of this paper.

More than half of the known roAp stars were discovered photometrically from SAAO. One of the basic reasons behind this is that the Sutherland site has stable and good sky transparency, allowing a closer match to scintillation noise than at many other observing sites used in other roAp surveys.  However, in the last ten years that we have been running the Nainital-Cape Survey, we have noticed a gradual increase in sky brightness and atmospheric noise owing to enhanced human activities around the ARIES and Sutherland observatories. It can be inferred from the scaling relation (eqn. \ref{scieq}) that the 
combined atmospheric noise can be minimized by installing bigger telescopes at a good observing site where 
one can find stable photometric sky conditions (Young \cite{youn67}). A new 1.3-m optical telescope is now operational at a new astronomical site of ARIES observatory known as Devasthal (longitude: $79^{o}40^{'}57{''} $ E, latitude : $29^{o}22{'}26{'}' $ N, altitude : 2420-m). 
In addition, a new 3.6-m telescope has been recently installed at the Devasthal site and is likely to be operational by 2016. The theoretical scintillation noise estimated for this telescope is 0.0217 mmag making the telescope very efficient for detecting tiny amplitude variations. 
The 0.5-m telescope of SAAO is also soon to be replaced with a 1.0-m robotic telescope. 
These up-coming  observing facilities equipped with modern state-of-the-art instruments at ARIES and SAAO will be the next step to boost  the Nainital-Cape survey and other projects aimed at the detection of sub-mmag light variations.

\section{New Null Results from the Nainital-Cape Survey}
\label{nulls}

In this paper we report the non-detections of pulsation in 108 CP stars. 
The first and second panels of Fig. \ref {null1} depict the light curves of the candidate stars observed from  ARIES and SAAO. The pre-whitened frequency spectra of the respective time-series are plotted in the third and fourth columns. The name of the star, duration of observations in hours and the heliocentric Julian dates are marked in each panel.

Here it is worth recalling  that roAp stars show amplitude 
modulation due to rotation and beating between multiple pulsation frequencies. Therefore, 
the non-detection of light variations  may be due to fact that the observations are acquired at a time when the pulsations are below the detection limit of the survey. For example, Joshi et al. (\cite{josh06}) classified HD\,25515 as a null result and then  subsequently, after further observations, classified it as a ${\delta}$ Scuti type pulsating variable (Joshi et al. \cite{josh09}).  
Hence, a null-detection of pulsations does not mean that the star is non-variable, but rather that its light output was not detected to vary during the particular interval of the observations. This demonstrates the necessity for repeated observations of the candidate stars. These null results are also an  important contribution toward understanding the distinction between 
pulsating and non-pulsating CP stars that are otherwise similar in all other observational respects (Murphy et al. \cite{murp15}). As mentioned above, a by-product of these null results is an observational characterisation of a particular observing site for data acquired on many nights over a wide range of observing conditions.

\section{Comparison of Known roAp Stars with the Null Results}
\label{comp}

At the time the Nainital-Cape survey began, only 23 roAp stars were known. Therefore, our knowledge of the extent of the roAp phenomenon at that time was used to define the target selection and observing strategy. Since then, the number of known roAp stars has more than doubled, and currently stands at 61 confirmed members of this class. The compilation of the various physical parameters of the known roAp stars are important to study the roAp and noAp (``non-roAp'') phenomena in Ap stars. Tables \ref{roap01} and \ref{roap02} list the  astrophysical parameters of the known roAp stars extracted from the available sources in the literature. For each star Table \ref{roap01} lists, respectively, the table entry  number, the HD number of roAp star, their popular name, spectral type, Str\"{o}mgren indices $b-y$,  $m_1$,  $c_1$, $\beta$, $\delta m_1$, $\delta c_1$, effective temperature $T_{eff}$, and the reference(s) from which the data were taken. Table \ref{roap02} lists the table entry number, the HD or HR catalogue number and other name(s)  of the roAp star, the visual magnitude $m_v$, parallax $\pi$ , distance d, absolute magnitude $M_v$, luminosity parameter log $(\frac{L_\star}{L_\odot})$, 
pulsational period corresponding to the highest amplitude, the frequency separation 
$\Delta \mu$, maximum photometric amplitude variation $A_{max}$, maximum 
radial velocity variation $RV_{max}$, rotational period $P_{rot}$, surface 
gravity log~$g$, mass $M_\star$, radius $R_\star$, mean longitudinal magnetic 
field and the projected rotational velocity $vsini$. Where no data is available in the data archives or in the literature for a given parameter, this is denoted with a  ``-''  symbol in the relevant column.
It is instructive to compare the coverage of the Nainital-Cape survey with the currently established extent of the roAp phenomenon. Therefore, the catalogue of the  basic parameters of the known roAp stars can be used for the statistical analysis of roAp and noAp phenomena in Ap stars located in the same part of the H-R diagram.

\section{Evolutionary States of the Studied Samples}

To establish the evolutionary status of the sample null result stars, we first established their luminosities and effective temperatures, which then allowed us to compare them with the known roAp stars. The absolute magnitudes and luminosities of the candidate  stars observed in the Nainital-Cape survey were determined based on the data 
taken from the Hipparcos catalogue (van Leeuwen \cite{vanl07}). The photometric $T_{eff}$ is calculated from the Str\"omgren $\beta$ indices using the grids of Moon \& Dworetsky (\cite{moon85}) that give a typical error of about 200 K. The various astrophysical parameters of the stars observed in the Nainital-Cape survey are listed in Table \ref{null1}.  These parameters are either taken from the Simbad database or calculated using the standard relations (Cox \cite{cox99}). For each star, this Table lists the HD number, right ascension $\alpha_{2000}$, 
declination $\delta_{2000}$, visual magnitude $m_v$, spectral type, parallax $\pi$, Str\"{o}mgren indices $b-y$,  $m_1$,  $c_1$, $\beta$, $\delta 
m_1$, $\delta c_1$, effective temperature $T_{eff}$, 
luminosity parameter log $(\frac{L_\star}{L_\odot})$, duration of the 
observations $\Delta t$, heliocentric Julian dates (HJD:2450000+) and year of 
observations (2000+) when the star was observed. The Str\"omgren indices $\delta m_1$ and  $\delta c_1$ are calculated using the calibration of Crawford (\cite{craw75}, \cite{craw79}).

The absolute magnitude $M_v$ in the $V$-band was determined using the 
standard relation (Cox \cite{cox99}),

\begin{equation}
M_v = m_v + 5 + 5\,log\, \pi -Av,
\end{equation}
\noindent where $\pi$ is trigonometric parallax measured in arcsec,
the interstellar extinction in the V-band is $A_V$ = $R_V$E(B-V)=3.1E(B-V). The 
 reddening parameter E(B-V) is obtained by taking the difference of the 
observed colour (taken from the Simbad data base) and intrinsic colour (estimated 
from Cox (\cite{cox99})).

The stellar luminosity was calculated using the relation
\begin{equation}
\log\dfrac{L}{L_{\sun}} = - \dfrac{M_V + BC - M_{{\rm bol},\sun}}{2.5},
\end{equation}
\noindent where we adopted the solar bolometric magnitude $M_{bol,\odot}$ = 4.74 mag (Cox 
\cite{cox99}), and  used the standard bolometric correction $BC$ from Flower (\cite{flow96}). Taking into account all 
contributions to the $M_v$ and $\frac{L_\star}{L_\odot}$ error budgets, we find 
a typical uncertainty of 20-25\% for both parameters. \\

\begin{figure}
\includegraphics[width=0.95\linewidth]{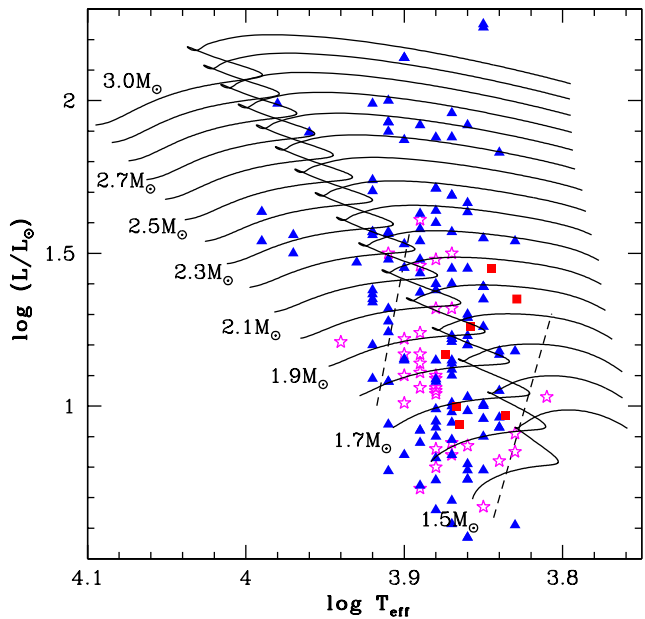}
\caption{The positions of the of null results (filled triangle) and $\delta$-Scuti type variables discovered under the Nainital-Cape survey (filled square). For comparison, 
positions of known roAp stars are also shown (open star). The solid 
lines show theoretical evolutionary tracks from the ZAMS (Christensen-Dalsgaard \cite{chris1993}). The dashed lines indicate the red and blue edges of the instability strip.}
\label{HR}
\end{figure}

The null objects displayed in Fig. \ref{HR} include all the objects from paper I,II,III and  IV (this paper) of the Nainital-Cape survey. The positions of known roAp stars and the newly discovered 
$\delta$-Scuti type variables in our survey are also shown. The evolutionary tracks for stellar masses ranging from 1.5 to 3.0  $M_\odot$ (Christensen-Dalsgaard \cite{chris1993}) are over-plotted. The position of the blue (left) and red (right) edges of the instability strip are shown by two oblique lines (Turcotte et al. \cite{turc00}). Fig. \ref{HR} clearly shows that most of the sample stars are located within  the instability strip. For reasons given above, we may expect that some of the stars listed as null results in this paper may well turn out to be variables in near future. However, with each subsequent non-detection of pulsations, the constraint on non-variability will be strengthened and they will be established as ``noAp'' stars, thus helping to shed light on the other observational characteristics that will allow us to distinguish between pulsating and constant CP stars, which is one of the long-term goals of the Nainital-Cape survey.

\section{Ground Based Surveys for Pulsation in Chemically Peculiar stars}
\label{survey}

In the past, several surveys have been conducted around the globe to search for roAp stars with different instrumental set-ups in both the northern and southern hemisphere independently. Such surveys required much telescope time, hence the photometric surveys were performed on 1-m class telescopes, on which it was possible to secure ample telescope time. Spectroscopic surveys became  more popular in recent years due to improved sensitivity of the high-resolution spectroscopic instruments used to search for low-amplitude oscillations in roAp candidates. The major drawback of this technique remains the small amount of observing time available on large telescopes. In this section,we provide a short description of the various surveys conducted for pulsation in CP stars.

\subsection{Cape Survey }

Following the discovery of first roAp star HD\,101065 in 1978, only 14 stars were known prior to 1990. A systematic survey for roAp stars in the southern hemisphere was initiated by Don Kurtz and Peter Martinez at SAAO with two objectives: the first was to increase the number of members of this class and the second was to study the relationship between the roAp stars and the other pulsating stars located at the same region of the H-R diagram. The observations for this survey were acquired with the photoelectric photometers attached to the 0.5-m and 1.0-m telescopes at Sutherland. Under the Cape-survey 134 Southern Ap SrCrEu stars were checked for the pulsational variability and 12 new roAp stars were discovered (Martinez et al. \cite{mart91}, Martinez \& Kurtz \cite{mart94a}, \cite {mart94b}).

\subsection{Nainital-Cape Survey}

The detection of the small amplitude light variations needs a lot of observational expertise. As mentioned above most of the roAp stars known prior to 2000 were discovered under the Cape survey, where  the SAAO astronomers gained a lot of observational experience. However, this meant that most of the known roAp stars were southern objects. The Nainital-Cape survey was initiated in 1999 as a collaboration between South African and Indian astronomers to increase the number of known roAp stars in the northern sky. This survey was started in 1999 and lasted for ten years making it the most extensive survey for pulsation in CP stars, where a total of of 337 Ap and Am stars were monitored. Although only one new roAp star, HD\,12098, was discovered, in this survey, it revealed milli-magnitude level light variations with periods similar to those of  the  $\delta$-Scuti stars in seven Am stars.  This survey is thus unique in a sense that both the Ap and Am stars were included in the samples, hence there were plenty of chances to discover pulsations in CP1 and CP2 stars. The null results of this survey have been published in Martinez et al. (\cite{mart01}), Joshi et al. (\cite{josh06,josh09}) as well as in the present paper. The archive of well-established null results is useful to delineate the extent of the roAp phenomenon and also to shed light on the distinction between roAp and noAp stars.

\subsection{Lowell-Wisconsin Survey}

Between 1985 to 1991, Nelson \& Kreidl (\cite{nels93}) conducted a survey of pulsation in 120 northern Ap stars of spectral range B8--F4. Though these authors did not report the discovery of any new roAp stars from their survey,  their main finding was the absence of pulsation in the spectral range B8--A5, indicating that roAp-like oscillations are likely to be confined to the cooler peculiar stars. 

\subsection{The Hvar Survey}

A photometric survey was initiated in 2011 to search for new northern roAp stars at the Hvar observatory (Paunzen et al. \cite{paun12}, \cite{paun15}). For this survey, a  CCD based photometer attached to the 1.0-m Austrian-Croatian telescope was used for  the observations of candidate stars. Under this survey, 80 candidate roAp stars were examined for a total duration of 100 hours. Differential CCD photometry was performed to detect the light variations in the sample Ap stars. The authors have not reported any positive detections and have presented the frequency spectra and the basic parameters of the null results they observed.

\subsection{Other Minor Photometric Surveys}

In addition to the above surveys, a number of smaller photometric surveys have also been conducted in the northern and southern hemisphere independently by  Dorokhova \& Dorokhov (\cite{doro98}), Kurtz (\cite{kurt82}), Matthews et al. (\cite{matt88}), Heller \& Kramer (\cite{hell90}), Schutt (\cite{schu91}), Belmonte (\cite{belm89}), Hildebrandt (\cite{hild92}), and Handler \& Paunzen (\cite{hand99}). Though these surveys are small in terms of sample size and  number of newly discovered roAp stars, they have helped to define candidate selection criteria for other roAp surveys.

\subsection{Spectroscopic Surveys}
Spectroscopy of high spectral and temporal resolution using  large telescopes allows the detailed study of line profile variations (Hatzes \& Mkrtichian \cite{hatz05}). After the discovery of significant RV pulsational variations in some known roAp stars (Kanaan \& Hatzes \cite{kana98}), candidate roAp stars have been monitored with time-resolved high-resolution spectroscopic observations by several observers in the last ten years. These observations have revealed that the highest RV amplitudes are observed in the spectral lines of the rare-earth elements, while spectral lines of the other elements show weak or undetectable oscillations. Using spectroscopic techniques, about 15 roAp stars have been discovered (Kochukhov \cite{koch06}; Kochukhov et al. \cite{koch08}; \cite{koch09}; Kochukhov et al. \cite{koch13}; Alentive et al. \cite{alen12}; Elkin et al. \cite{elki05a}; \cite{elki05b}; Kurtz et al. \cite{kurt06}).

\section{Conclusions}
\label{conc}
In this paper, we have presented the light curves and frequency spectra of the 108 candidate stars observed in the Nainital-Cape survey. 
Analyses of the photometry acquired at Sutherland and Nainital indicate that we have achieved a detection level of about 0.6 mmag in the frequency range 1-5 mHz in the Nainital-Cape survey.  Using the standard  relations and data extracted from the literature we have presented the various  astrophysical parameters of the null results. We have also compiled the basic physical parameters of the known roAp stars. On comparing  the positions of the known roAp stars to the observed sample stars in the H-R diagram, we infer that  the boundary of the roAp phenomenon extends beyond the cool edge of the classical instability strip.

\section*{Acknowledgments}
This work was carried out under the Indo-South Africa Science and Technology Cooperation INT/SAFR/P-3(3)2009) and NRF grant UID69828 funded by Departments of Science and Technology of the Indian and South African Governments. SC acknowledges support under the Indo-Russian grant INT/RFBR/P-118 through which he received a stipend to perform this work. SC also thanks the ARIES academic committee for their help and support. We acknowledge use of SIMBAD, NASA's ADS and ESA's Hipparcos database.

\small

\onecolumn
\begin{landscape}
\begin{longtable}{crrccccccrrccc}
\caption{The known roAp stars and their physical parameters.}\label{roap01}\\
\hline\hline
{S.N.}&{Star Name}&Other&{$\alpha_{2000}$}&{$\delta_{2000}$}&{Sp.}&{b-y}&{$m_{1}$}&{$c_{1}$}&{$\delta m_{1}$}&{$\delta c_{1}$}&{$\beta$}&{$T_{eff}$} & Ref.\\
   &  & Name   &          &         & Type            & mag   &  mag  &  mag  &  mag &  mag & mag  & K &      \\
\hline
\endfirsthead
\hline
\caption{continued.}\\
\hline\hline
{S.N.}&{Star Name}&Other&{$\alpha_{2000}$}&{$\delta_{2000}$}&{Sp.}&{b-y}&{$m_{1}$}&{$c_{1}$}&{$\delta m_{1}$}&{$\delta c_{1}$}&{$\beta$}&{$T_{eff}$} & Ref.\\
   &   & Name   &          &          & Type            & mag   &  mag  &  mag  &  mag &  mag & mag  & K & \\
\hline
\endhead
\endfoot
\footnotetext{1. Holdsworth et al. (\cite{hold14a}); 2. Kurtz et al. (\cite{kurt96}); 3. Kurtz et al. (\cite{kurt94b}); 4. Martinez \& Kurtz (\cite{mart94a}); 5. Girish et al. (\cite{giri01}); 6. Schneider et al. (\cite{schn92}); 7. Kurtz (\cite{kurt82}); 8. Kurtz et al. (\cite{kurt05b}); 9. Mkrtichian \& Hatzes (\cite{mkrt05a}); 10. Matthews et al. (\cite{matt87}); 11. Elkin et al. (\cite{elki11}); 12. Kochukhov et al (\cite{koch09}); 13. Kurtz et al. (\cite{kurt97b}); 14. Kochukhov (\cite{koch06}); 15. Kurtz et al. (\cite{kurt97a}); 16. Kurtz \& Martinez (\cite{kurt94c});  17. Elkin et al. (\cite{elki10}); 18. Dorokhova \& Dorokhov (\cite{doro98}); 19. Handler et al. (\cite{hand06}); 20. Elkin et al. (\cite{elki05b}); 21. Mkrtichian \& Hatzes (\cite{mkrt05b}) 22. Martinez \& Kurtz (\cite{mart90a}); 23. Kochukhov et al (\cite{koch08}); 24. Elkin et al. (\cite{elki05a}); 25. Martinez et al. (\cite{mart93}); 26. Handler et al. (\cite{hand02}); 27. Balona \& Laney (\cite{balo03}); 28. Kurtz et al. (\cite{kurt94a}); 29. Kochukhov et al. (\cite{koch13}); 30. Kreidl \& Kurtz (\cite{krei86}); 31. Hatzes \& Mkrtichian (\cite{hatz04}); 32. Kurtz et al. (\cite{kurt05a}); 33. Mkrtichian et al. (\cite{mkrt03}); 34. Kurtz (\cite{kurt91}); 35. Kurtz et al. (\cite{kurt06}); 36. Martinez et al. (\cite{mart91}); 37. Kurtz et al. (\cite{kurt03}); 38. Kurtz \& Martinez (\cite{kurt87}); 39. Holdsworth et al. (\cite{hold14b});
40. Hatzes \& Mkrtichian (\cite{hatz05}); 41. Heller \& Kramer (\cite{hell90}); 42. Kurtz et al. (\cite{kurt11}); 43. Balona et al. (\cite{balo13}); 44. Alentiev et al. (\cite{alen12}); 45. Holdsworth (\cite{hold15}); 46. Kurtz \& Martinez (\cite{kurt95}); 47. Balona et al. (\cite{balo11}); 48. Smalley et al.  (\cite{smal15}); 49. Kochukhov \& Ryabchikova (\cite{koch01}); 50. Martinez et al. (\cite{mart96}); 51. Martinez et al. (\cite{mart90b}); 52. Martinez et al. (\cite{mart98}); 53. Kreidl et al. (\cite{krei91}); 54. Gonzalez et. al (\cite{gonz08}).}
1. &         & J0008         &00 08 30 & +04 28 18&A9p SrEu(Cr)&-&     -&     -&       -&      -&-   & 7300&1   \\
2. &   HD6532 & &01 05 56 &-26 43 44  & ApSrCrEu    &0.082 & 0.233&0.870 & -0.014& -0.051 & 2.882&7900 & 2      \\ 
3. &   HD9289 & &01 31 16 &-11 07 08  & ApSrEu      &0.138 &0.225 &0.826 &-0.018  &-0.012 &2.833 &8000 & 3,4  \\
4. &  HD12098 & &02 00 40 &+58 31 37  & F0        &0.191 &0.328 &0.517 &-0.122  &-0.279 &2.796 &7500 & 5     \\
5. &  HD12932 & &02 06 16 &-19 07 26  & ApSrEuCr    &0.179 &0.228 &0.765 &-0.024  &-0.035 &2.810 & 7620& 6    \\
6. &  HD19918 & &03 00 37 &-81 54 07  & Ap SrEuCr  &0.169 &0.216 &0.822 &-0.010  &-0.058 &2.855 &7800 & 4    \\
7. &  HD24355 &J0353     &03 53 23 & 25 38 33 & A5p SrEu  & -    & -    &    -   & -     & -     &  -   &8250 & 1   \\
8. &  HD24712& HR\,1217 &03 55 16&-12 05 57  & ApSrEu(Cr)&0.191 &0.211 &0.626   &-0.023 &-0.074 &2.744 &7250 & 7,8,9 \\
9. &  HD42659 & &06 11 22 &-15 47 35 & ApSrCrEu      &0.124 &0.257 &0.765 &-0.050  &-0.076 &2.834 &7500 & 4      \\
10.&  HD258048& J0629         &06 29 57 & +32 24 47&F4p EuCr(Sr)&-&     -&     -&       -&      -&- & 6600&1   \\
11.&          & J0651         &06 51 42 & -63 25 50&F0p SrEu(Cr)&-&     -&     -&       -&      -&- & 7400&1   \\
12.&  HD60435 & &07 30 57 &-57 59 28  & ApSr(Eu)    &0.136 &0.240 &0.833 &-0.034  &-0.047 &2.855 &7800 &10    \\
13.&  HD69013 & &08 14 29 &-15 46 32 & A2SrEu          &0.296 &0.330 &0.400 &-0.137  &-0.324&2.772 &7600 & 11    \\
14.&  HD75445&  &08 48 43 &-39 14 02  & SrEu        &0.159 &0.218 &0.729 &-0.019  &0.001 &2.801 &7650& 12     \\
15.&          & J0855         &08 55 22 & +32 42 36&A6p SrEu    &-&     -&     -&       -&      -&- & 7800&1   \\
16.&  HD80316 & &09 18 25 &-20 22 16  & ApSr(Eu)    &0.118 &0.324 &0.599 &-0.118  &-0.283 &2.856 &7700& 13       \\
17.&  HD83368 & HR\,3831 & 09 36 25 &-48 45 04 & A8pSrEuCr &0.159 &0.230 &0.766 &-0.024  &-0.062 &2.825 &7650& 7,14,15  \\
18.&  HD84041 & &09 41 34 &-29 22 25  & ApSrEuCr    &0.177 &0.233 &0.797 &-0.026  &-0.061 &2.844 &7500& 4       \\
19.&  HD86181 & &09 54 53 &-58 41 45 & ApSr        &0.172 &0.205 &0.757 & 0.001  &-0.061 &2.819 &7900& 16     \\
20.&  HD92499 & &10 40 08 &-43 04 51  & A2SrCrEu    &0.179 &0.301 &0.615 &-0.099  &0.000 &2.812 &7500& 17       \\ 
21.&  HD96237 & &11 05 34 &-25 01 09 & A4SrEuCr           &0.233 &0.261 &0.704 &-0.054  &-0.122 &2.824 &7800& 11      \\
22.& HD97127  & J1110         &11 10 54 & +17 03 48&F3p SrEu(Cr) &-&     -&     -&       -&      -&- & 6300&1   \\
23.&  HD99563 & &11 27 17 &-08 52 08  & F0        &0.171 &0.206 &0.745 &-0.001  &-0.090 &2.830 &7000& 18,19,20    \\
24.& HD101065 & Przbylski's star &11 37 37 &-46 43 00 & Controversial &0.431 &0.387 &0.002 &-0.204  &-0.370 &2.641 &6800& 7,21,22 \\
25.& HD115226 & &13 18 00 &-72 57 01 & A3pSr       &     -&     -&     -&      - &     - &    - &7600& 23      \\
26.& HD116114 & &13 21 46 &-18 44 32 & ApSrCrEu   &0.172 &0.226 &0.843 &-0.016  & 0.008 &2.836 &7600& 24      \\
27.& HD119027 & &13 41 20 &-28 46 60  & ApSrEu(Cr) &0.257 &0.214 &0.557 &-0.034  &-0.076 &2.731 &7050& 25      \\
28.&          & J1430    &14 30 50    & +31 47 55 &A9p SrEu    &-&     -&     -&       -&      -&- & 7100&1  \\
29.& HD122970 & &14 04 49 &+05 24 51  & F0  &0.260 &0.178 &0.540 &-0.005  &-0.011 &2.707 &7000& 26      \\
30.& HD128898 & $\alpha$ Cir &14 42 30 &-64 58 30 & ApSrEu(Cr)&0.152 &0.195 &0.760 & 0.012  &-0.077 &2.831 &7500& 7,27,28       \\
31.& HD132205 && 15 00 04 & -55 02 60 & A2EuSrCr &-&-&-&-&-&-& 7800 &29 \\
32.& HD134214 & &15 09 02 &-13 59 59& ApSrEu(Cr)  &0.216 &0.223 &0.620 &-0.029  &-0.108 &2.774 &7400& 30      \\
33.& HD137909 & $\beta$ CrB & 15 27 50 &+29 06 20& F0p  &0.141 &0.257 &0.740 &-0.056 &0.002 &2.839 &7800& 31 \\
34.& HD137949 & 33 Lib & 15 29 35 &-17 26 27& ApSrEuCr&0.196 &0.311 &0.580 &-0.105&-0.236 &2.818 &7700&32,33,34 \\
35.& HD143487 & &16 01 44 &-30 54 57& A3SrEuCr  &0.311 &0.262 &0.393 &-0.089  &-0.169 &2.706 &7000& 17       \\\
36.& HD148593 & & 16 29 39 & -14 35 06 & A2 Sr &- & -&- & -&- & -& 7850 & 29 \\
37.&          & J1640         &16 40 03 & -07 37 30&A8p SrEu  &-&     -&     -&       -&      -&- & 7400&1  \\
38.& HD150562&  &16 44 11 &-48 39 18 & A/F (pEu)   &0.301 &0.212 &0.659 &-0.015  &-0.087 &2.783 &7500& 4    \\
39.& HD151860 & & 16 52 59 & -54 09 46  & A2SrEu & 0.327 & 0.221 & 0.538 & -&-  & - & 7050 & 29 \\
40.& HD154708 & &17 10 28 &-58 00 17 & Ap    &0.277 &0.256 &0.464 &-0.079  &0.015 &2.757 &7200& 35      \\
41.& HD161459 & &17 48 30 &+51 55 02 & ApEuSrCr    &0.245 &0.246 &0.679 &-0.040  &-0.141 &2.820 &7950& 36      \\ 
42.& HD166473 & &18 12 26 &-37 45 09 & ApSrEuCr    &0.208 &0.321 &0.514 &-0.118  &-0.268 &2.801 &7700& 37,38      \\
43.& &KIC\,007582608          &18 44 12 & +43 17 51 & Ap      &-&     -&       -&      -& - &  - &8700  &39  \\
44.& HD176232 & 10 Aql &18 58 47 &+13 54 24 & F0pSrEu &0.150 &0.208 &0.829 &-0.004  & 0.031 &2.809 &7400& 40,41 \\
45.& &KIC\,010195926&19 05 27&+47 15 48&          Ap&     -&     -&     -&       -&      -&     -&7400& 42  \\
46.& &KIC\,008677585&19 06 28&+44 50 33&A5p         &     -&     -&     -&       -&      -&     -&7600& 43   \\
47.& HD177765 & & 19 07 10 &  -26 19 54 & A5SrEuCr & 0.248 &0.261 &0.731 &-0.054&-0.110 &2.834& 8000 & 44\\
48.&          & J1921         &19 21 29&+47 10 53 & F3p SrEuCr & -    &-     &     -&     -&       -&   - & 6200&45   \\
49.& HD185256 &               &19 39 20 &-29 44 34 & ApSr(EuCr) &0.277 &0.185 &0.615 &-0.004& -0.039 &2.738 &7250& 46 \\
50.&          & J1940         &19 40 08 &-44 20 09 &F2(p Cr)    &-    &     -&     -&      -&      -&- & 6900&1  \\
51.& &KIC\,010483436&19 46 29&+47 37 50&          Ap&     -&     -&     -&       -&      -&     -&7388& 47 \\
52.& HD225914 & KIC\,004768731&19 48 26 & +39 51 58& Ap&     -&     -&     -&       -&      -&- & 7726 & 48  \\
53.& HD190290 & &20 13 56 &-78 52 42 & ApEuSr      &0.289 &0.293 &0.466 &-0.091  &-0.306 &2.796 &7500& 36      \\
54.& HD193756 & &20 24 12 &-51 43 25 & ApSrCrEu    &0.181 &0.213 &0.760 &-0.008  &-0.040 &2.810 &7500& 36      \\
55.& HD196470 & &20 38 10 &-17 30 06  & ApSrEu(Cr)   &0.211 &0.263 &0.650 &-0.059  &-0.144 &2.807 &7850& 36     \\
56.& HD201601 & $\gamma$ Equ &21 10 20 &+10 07 54 & F0p &0.147 &0.238 &0.760 &-0.032  &-0.058 &2.819 &7600& 49,50 \\
57.& HD203932 & &21 26 04 &-29 55 48 & ApSrEu     &0.175 &0.196 &0.742 & 0.004  &-0.020 &2.791 &7200& 51     \\
58.& HD213637 & &22 33 12 &-20 02 22 & A(pEuSrCr)    &0.298 &0.206 &0.411 & -0.035  &-0.031 &2.670 &6400& 52      \\
59.& HD217522 & &23 01 47 &-44 50 27 & Ap(Si)Cr  &0.289 &0.227 &0.484 &-0.056  &-0.015 &2.691 &7100& 53    \\
60.& HD218495 & &23 09 28 &-63 39 12 & A2pEuSr     &0.114 &0.252 &0.812 &-0.049  &-0.098 &2.870 &8000& 36       \\
61.& HD218994 & &23 13 16 &-60 35 03 & A3Sr        &0.154 &0.196 &0.826 & 0.008  & 0.032 &2.807 &7600& 54    \\
\hline
\end{longtable}
\end{landscape}
\begin{landscape}
\begin{longtable}{crrrrrrrrrrrcccccc}
\caption{Additional parameters for the known roAp stars.}\label{roap02}\\
\hline\hline
S.N.& Star Name &{$m_{v}$}&{$\pi$}&{d}&{$M_{V}$}&{log$(L_{\star}/L_{\odot})$}&{$P_{pul}$}&{$\Delta\mu$}&{$A_{max}$}&{$RV_{max}$}&{$P_{rot}$}&{log g}&$M_\star$ &$R_{\star}$&  Mag. Field  &  vsini \\
& & mag    & mas   & pc & mag &                    & min       &  $\mu$Hz         &  mmag           & kms$^{-1}$      &  days     & dex  & $M_{\odot}$ & $R_{\odot}$ & kG & kms$^{-1}$ \\ \hline
\hline
\endfirsthead
\caption{continued.}\\
\hline\hline
S.N.& Star Name &{$m_{v}$}&{$\pi$}&{d}&{$M_{V}$}&{log$(L_{\star}/L_{\odot})$}&{$P_{pul}$}&{$\Delta\mu$}&{$A_{max}$}&{$RV_{max}$}&{$P_{rot}$}&{log g}&$M_\star$ &$R_{\star}$&  Mag. Field  &  vsini \\
& & mag    & mas   & pc & mag &                    & min       &  $\mu$Hz         &  mmag           & kms$^{-1}$      &  days     & dex  & $M_{\odot}$ & $R_{\odot}$ & kG & kms$^{-1}$ \\ \hline
\endhead
\hline
\endfoot
1. &   J0008   &  10.16&  -  &   -   &  -  &  -  & 9.58& - & 0.76 &  -  &  -  &  -  &-   &   -&    -    &   -  \\
2. &   HD6532  &  8.40 & 6.14&162.87 &2.20 &1.22 & 7.10&47 & 5.00 &1.15 &1.94  &4.30 &-   &   -&   0.22  & 30     \\
3. &   HD9289  &  9.38 &    -&     - &2.42 &    -&10.52& - & 3.50 &0.85 &8.55  &4.15 &-   &   -&    0.65 &10.5      \\
4. &  HD12098  &  8.07 &    -&     - &   - &    -& 7.61& - & 3.00 &   - &5.46  &4.20 &1.70&1.70&   1.46 & 10     \\
5. &  HD12932  & 10.25 &    -&     - &2.55 &    -&11.61& - & 4.00 &1.40 &3.53  &4.15 &-   &   -&   1.20 &2.50      \\
6. &  HD19918  &  9.34 & 4.07&245.70 &2.34 &1.06 &11.04& - & 2.00 &1.30 & -    &4.34 &-   &   -&   1.60 &3.00      \\
7. &  HD24355  &  9.65 &  -  &   -   &  -  &  -  & 6.42& - & 1.38 &  -  &13.95 &  -  &-   &   -&    -    &   -  \\
8. &  HD24712  &  6.00 &20.32& 49.21 &2.32 &0.87 & 6.13&68 &10.00 &0.25 &12.46 &4.30 &1.55&1.77&3.10 &5.60  \\
9. &  HD42659  &  6.76 & 7.60&131.58 &2.38 &1.48 & 9.70&52 &0.80  &0.70 &    - &4.40 &2.10&   -&0.39 & 19.00  \\
10.&  HD258048 & 10.52 &  -  &   -   &  -  &  -  & 8.49& - &1.49  &  -  &  -  &  -  &-   &   -&    -    &   -  \\
11.&   J0651   & 11.51 &  -  &   -   &  -  &  -  &10.88& - &0.79  &  -  &  -  &  -  &-   &   -&    -    &   -  \\
12.&  HD60435  &  8.89 & 4.41&226.76 &1.54 &1.14 &11.90&52 &16.00 &1.90 &7.68  &4.40 &1.82&   -&0.30  &10.8 \\
13.&  HD69013  &  9.56 &    -&      -&    -&    -&11.22& - & -    &0.20 &-     &4.50 &-   &   -&2.90  &6.0 \\
14.&  HD75445  & 7.12  & 9.30&108.34 &1.96 &1.17 & 9.00& - & -    &0.29 &2.08  &4.32 &1.81&  - &2.98                & 2\\
15.&   J0855   &10.80  &  -  &   -   &  -  &  -  & 7.30& - &1.40  &  -  &3.09  &  -  &-   &   -&    -    &   -  \\
16.&  HD80316  & 7.78  & 7.25&137.93 &2.26 &1.11 & 7.40& - & 2.00 &0.32 &2.08  &4.58 &1.70&1.53&0.18  & 32.0 \\
17.&  HD83368  & 6.17  &14.16& 70.62 &2.47 &1.09 &11.60& - &10.00 &3.33 &2.85  &4.20 &1.76&2.00&0.50  & 33.0 \\
18.&  HD84041  & 9.33  &    -&   -   &2.38 &   - &15   &60 & 6.00 &0.50 &3.69  &4.30 &   -&   -&0.48  & 25.0\\
19.&  HD86181  & 9.32  & 3.49&286.53 &2.49 &1.01 & 6.20& - & 4.60 &   - &   -  &-    &   -&   -&  0.40 & -\\
20.&  HD92499  & 8.89  & 3.54&282.48 &1.63 &1.05 &10.40& - &   -  &0.066&   -  &4.00 &1.68&   -&   8.15&3.3 \\
21.&  HD96237  & 9.43  & 1.53&653.59 &-    &1.61 &13.89& - &   -  &0.10 &   -  & 4.30&   -&   -&  2.90 &6 \\
22.&  HD97127  & 9.43  &  -  &   -   &  -  &  -  &13.51& - &0.66  &  -  &  -  &  -  &-   &   -&   -   &   -  \\
23.&  HD99563  & 8.67 & 3.92&255.10 &1.90 &1.10 &10.70& - &10.00 &4.9   &2.91  &4.20 &2.03&1.90&   0.57              &28.0 \\
24.& HD101065  & 7.99 & 8.93&111.98 &2.09 &0.91 &12.16&68 &13.00 & 1.03 &3.94  &4.20 &1.52&1.98   &2.30 & 4.0 \\
25.& HD115226  & 8.51 & 6.80&147.06 &2.67 &0.86 &10.86& - &   -  & 1.24 &3.30  &4.00 &1.60&  - &0.75 & 27  \\
26.& HD116114  & 7.02 & 7.71&129.70 &1.35 &1.32 &21.30& - &   -  & 0.65 &   -  &4.10 &2.07&  - &0.50 & 2.2\\
27.& HD119027  &10.02 &    -&     - &3.04 &0.67 & 8.63&52 &2.00  & 0.148&   -  &4.40 &   -&  - &3.10 & 4.0 \\
28.&   J1430   &11.56 &  -  &   -   &  -  &  -  & 6.11& - &1.06  &  -  &  -  &  -  &-   &   -&    -    &   -  \\
29.& HD122970  & 8.33 & 8.67&115.34 &2.94 &0.82 &11.18&68 &2.00  & 1.05 &3.88  &4.20 &1.50&1.80&0.22& 4.2     \\
30.& HD128898  & 3.20 &60.35&16.57  &1.90 &1.04 & 6.82&50 &5.00  & 0.80 &4.48  &4.20 &1.70&1.90&1.50  & 13.5  \\
31.& HD132205  & 8.72 &-    &      -&    -&    -& 7.14& - &  -  & 0.097& -    & 4.40 & -&- & 5.20& 9.50  \\ 
32.& HD134214  & 7.46 & 9.74&102.67 &2.60 &0.88 &5.69 & - &7.00  & 0.72&248    &4.05 &1.60&1.80&   2.70 & 2.6     \\
33.& HD137909  & 3.68 &29.17&34.28  &1.17 &1.46 &16.20& - &-     &0.04 &18.49  &4.40 &1.60&1.45&5.30 & 3.5   \\
34.& HD137949  & 6.67 &11.28&88.65  &1.88 &1.17 & 8.27&40 &3.00  & 0.33 &     -&4.30 &1.78&2.60& 4.70 & 3.0  \\
35.& HD143487  & 9.42 &    -&    -  &   - &    -& 9.63&-  &-     & 0.047&     -&5.00 &   -&  - & 4.70 & 1.5      \\
36.& HD148593  & 9.13 & -   & -     & -   & -   &10.69& - & -    &  -  & - & 4.40&-   & -  &  3.00& 5.00 \\
37.&   J1640   &12.67 &  -  &   -   &  -  &  -  &9.48 & - &3.52  &  -  &3.67  &  -  &-   &   -&  -  &   -  \\
38.& HD150562  & 9.82 &    -&    -  &2.68 &    -&10.80&50 &0.80  &  0.14&     -&4.40 &   -&  - &   5.00  & 1.5  \\
39.& HD151860  & 9.01 &   - &-      & -   &-    &12.30& - &-     &  -  &0.083 & 4.50 & - &- & 2.50 & 4.5  \\
40.& HD154708  & 8.76 & 6.75&148.15 &2.39 &0.73 & 8.00& - &-     &  0.09& 5.37 &4.11 &1.50&1.70&24.50 & 4.0     \\
41.& HD161459  &10.33 &    -&-      &2.47 &-    &12.00&-  &1.30  &     -&     -&4.38 &   -&   -&   1.76 & -     \\
42.& HD166473  & 7.92 &    -&-      &2.52 &1.24 & 8.80&68 &2.00  & 0.10 &     -&4.47 &1.80&   -&   8.50  & 2.5  \\
43.& KIC\,007582608&11.25& -&      -&   - & 1.21& 7.90&-  &1.45  & -    &20.45 &4.30 &2.37&1.77&  3.05          &-  \\
44.& HD176232  & 5.89 &12.76&78.37  &2.55 &1.32 &11.60&51 &0.60  & 0.54 &     -&4.10 &2.00&2.50&1.40 & 2.7     \\
45.& KIC010195926&10.66 &  -&   -   &   - &1.50 &17.14&55 &0.078 & 171  &5.68  &3.60 &1.70&3.60&5    &21  \\
46.& KIC008677585&10.19 &  -&   -   &   - &0.80 &10.28&37.2&0.033& -    &4.30  &3.90 &1.80&2.50&3.20 &4.2  \\
47.& HD177765  & 9.15 &    -&-      &-    & 1.50&23.6 & - &-    &0.148 &     -&3.80 & 2.20&-   & 3.60 &2.5      \\
48.&   J1921   &12.16 &  -  &   -   &  -  &  -  &11.18& - &1.99  &  -  &  -  &  -  &-   &   -&    -    &   -  \\
49.& HD185256  & 9.94 &    -&-      &-    &-    &10.33& - &3.00  & 0.15 &     -&4.30 &-   &-   & 0.71  &6.2     \\
50.&   J1940   &13.02 &  -  &   -   &  -  &  -  & 8.16& - &4.16  &  -  & 9.58 &  -  &-   &   -&    -    &   -  \\
51.& KIC\,010483436&11.43& -&    -  &   - &0.84 &12.32&-  &0.068 & -    &4.30  &4.15 &1.60&1.61&    -   & 20  \\
52.& KIC\,004768731&9.17 & -&    -  &   - &-    & 23.4&-  &0.062 & -    & -    &-    &-   &-   &   -    &-  \\
53.& HD190290  & 9.91 &    -&-      &2.49 &-    & 7.34&40 &2.00  & 0.50 &4.03  &4.54 &-   &-   &3.23  & 16     \\
54.& HD193756  & 9.20 &    -&-      &2.55 &-    &13.00&-  &0.90  & 0.74 &     -&4.29 &-   &-   &0.19  & 17.0  \\
55.& HD196470  & 9.72 &    -&-      &2.52 &-    &10.80&-  &0.70  & -    &     -&4.37 &-   &-   &1.48  & -  \\
56.& HD201601  & 4.68 &27.55&36.30  &2.49 &1.10 &12.40&30 &3.00  & 0.58 &     -&4.20 &1.74&2.16&3.80& 2.5     \\
57.& HD203932  & 8.82 &    -&-      &2.65 &-    &5.94&66  &2.00  & 0.33 &-&4.30&-    &-   &  0.267 & 4.7   \\
58.& HD213637  & 9.61 &    -&-      &-    &1.03 &11.50&-  &1.50  & 0.36 &$<$25 &3.60 &1.60 &-   &0.74 &  3.5  \\
59.& HD217522  & 7.52 &11.36&88.03  &2.77 &0.85 &13.70&58 &4.00  & 0.12 &8.55  &4.20 &1.49&1.86&1.70 & 2.7     \\
60.& HD218495  & 9.38 &    -&-      &2.23 &-    & 7.44&-  &1.00  & 0.79 &-     &4.40 &-   &-   &0.91  & 16.0    \\
61.& HD218994  & 8.56 & 3.55&281.69 &1.25 &1.06 &14.20&-  &-     & 0.093&-     &4.10 &-   &-   & 0.440   &5.2     \\
\end{longtable}
\end{landscape}

\begin{landscape}
\begin{longtable}{rccccccccccccccccc}
\caption{CP stars observed for pulsation from ARIES and SAAO and classified as null results in this survey. Their physical parameters are listed.}
\\
\hline\hline
S.N. &{Star}&{$\alpha_{2000}$}&{$\delta_{2000}$}&{$m_{v}$}&{Sp. }&{$\pi$}&{b-y}&{$m_{1}$}&{$c_{1}$}&{$\beta$}&{$\delta m_{1}$}&{$\delta c_{1}$}&{$T_{eff}$} &{log($L/L_{\odot}$)} &{$\Delta$t}&{HJD}&{Year of} \\
& HD   &                    & &mag & Type & mas            & mag   &  mag  &  mag  &  mag &  mag & mag   & K     &  & hr &  &Observation    \\
\hline
\endfirsthead
\caption{continued.}\\
\hline\hline
S.N. &{Star}&{$\alpha_{2000}$}&{$\delta_{2000}$}&{$m_{v}$}&{Sp. }&{$\pi$}&{b-y}&{$m_{1}$}&{$c_{1}$}&{$\beta$}&{$\delta m_{1}$}&{$\delta c_{1}$}&{$T_{eff}$} &{log($L/L_{\odot}$)} &{$\Delta$t}&{HJD}&{Year of} \\
& HD   &                    & &mag & Type & mas            & mag   &  mag  &  mag  &  mag &  mag & mag   & K     & & hr &  & Observation \\
\hline
\endhead
\hline
\endfoot
1.&   1169 &00 16 05&+08 06 56& 7.60&A5   & 8.83$\pm$0.69& 0.187& 0.240& 0.716& 2.772&-0.047&-0.008& 7455&1.18 & 0.45 & 4365 & 07 \\
&  &   &&&& & & & & & && & & 1.07 & 4366&07 \\
2.&  1486 & 00 19 18 & +59 08 20 & 7.28 & B9V & 6.32$\pm$0.83 &  - &  - &  - &  - &  - &  - & - & -  & 2.78 & 4071 &06  \\
 &  &  &  &  &  &  &  &  &  &  &  &  &    &  & 2.41 & 4367 & 07 \\
 &  &  &  &  &  &  &  &  &  &  &  &  &   &  & 1.91 & 4375 & 07 \\
 &  &  &  &  &  &  &  &  &  &  &  &  &   &  & 3.73 & 4376 & 07 \\
 &  &  &  &  &  &  &  &  &  &  &  &  &   &  & 1.42 & 4400 & 07 \\
 &  &  &  &  &  &  &  &  &  &  &  &  &   &  & 2.18 & 4401 & 07 \\
 &  &  &  &  &  &  &  &  &  &  &  &  &   &  & 3.22 & 4402 & 07 \\
 &  &  &  &  &  &  &  &  &  &  &  &  &   &  & 2.84 & 4428 & 07 \\
 &  &  &  &  &  &  &  &  &  &  &  &  &   &  & 2.09 & 4429 & 07 \\
3.&   2837 &00 32 09&+43 42 42& 9.16&A0p  & 2.75$\pm$1.08& -    & -    & -    & -    & -    & -    & -   &-& 1.42 & 4459&07    \\
4.&  3321 & 00 36 22 & +33 38 39 & 8.42 & A3 & 6.34$\pm$0.90 &  - &  - &  - &  - &  - &  - &- & - & 2.07 & 4397 & 07 \\
5.&   6757 &01 08 53&+45 12 27& 7.70&A0Vp & 3.35$\pm$0.91& -    & -    & -    & -    & -    & -    & -    &- &1.35& 4427&07      \\
&  &   &&&& & & & & & && &  &1.01 & 4431&07 \\
6.&   7676  &01 16 07&-34 08 56&8.37 &A5p   & 3.50$\pm$0.74&0.085 &0.280 &0.715 &2.830 &-0.073&-0.120&8008 &1.47&1.95 &4097 &06\\
7.&   8441 &01 24 19&+43 08 32& 6.67&A2p  & 4.88$\pm$0.59& 0.022& 0.141& 1.145& 2.833& 0.066& 0.306& 9617&1.97 &2.30 & 4751&08 \\
8.&   8783  &01 24 00&-72 19 28&7.82 &Ap    & 3.99$\pm$0.44&0.072 &0.199 &1.086 &-     &-     &-     &-    &-   &2.89 &4077 &06\\
9.&  11090 & 01 46 35 & -67 28 06 & 10.78 & ~ Ap &  &  - &  - &  - &  - &  - &  - &-& -  & 1.19 & 2127 & 01 \\
 &  &  &  &  &  &  &  &  &  &  &  &  &   &  & 3.45 & 2128 & 01 \\
10.&  11948 &01 58 51&+55 34 54& 7.85&F0p  & 6.71$\pm$0.73& 0.115& 0.242& 0.879& 2.873&-0.016& 0.027& 8323&1.50 &0.97 & 4459&07 \\
11.&  12211 & 02 00 33 & +27 53 19 & 9 & A7V & 7.12$\pm$1.97 &  - &  - &  - &  - &  - &  - & -& -  & 0.69 & 4399 & 07 \\
 &  &  &  &  &  &  &  &  &  &  &  &  &   &  & 4.42 & 4401 & 07 \\
 &  &  &  &  &  &  &  &  &  &  &  &  &   &  & 2.14 & 4402 & 07 \\
12.&  14433 & 02 21 55 & +57 14 34 & 6.39 & A1Ia & 0.79$\pm$0.46 & 0.463 & -0.088 & 0.913 & 2.606 &  - &  - &- & -  & 1.98 & 2238 & 01 \\
13.&  15144 & 02 26 00 & -15 20 28 & 5.86 & A6Vsp & 12.98$\pm$0.74 & 0.402 & 0.213 & 0.298 & 2.584 &  - &  - & - & -  & 1.86 & 4085 & 06 \\
14.&  15550 & 02 30 38 & +19 51 19 & 6.14 & A9V & 15.14$\pm$0.46 & 0.156 & 0.187 & 0.835 & 2.776 &  - &  - & -& -  & 1.54 & 4431 & 07 \\
15.&  16145 & 02 35 04 & -17 17 22 & 7.64 & Ap & 4.33$\pm$0.71 & 0.028 & 0.201 & 1.057 &  - &  - &  - & -& -   & 1.41 & 4087 & 06 \\
 &  &  &  &  &  &  &  &  &  &  &  &  &   &  & 1.99 & 4088 & 06 \\
 &  &  &  &  &  &  &  &  &  &  &  &  &   &  & 1.92 & 4089 & 06 \\
 &  &  &  &  &  &  &  &  &  &  &  &  &   &  & 1.92 & 4090 & 06 \\
16.&  17034 & 02 45 42 & +48 08 37 & 8.63 & B8V+ & 0.92$\pm$0.90 &  - &  - &  - &  - &  - &  - & -& -  & 1.68 & 4427 & 07 \\
17.&  17835 & 02 51 52 & +02 54 49 & 8.9 & ~ A4 & - & 0.16 & 0.17 & 0.96 & 2.84 &  - &  - &-& -  & 0.91 & 2216 & 01 \\
18.&  18078 &02 56 32&+56 10 41& 8.30&A0p  & -            & 0.087& 0.251& 1.079& 2.831&-0.044& 0.243& 7947  &- &0.96 & 4459&07    \\
19.&  18610  &02 54 18&-73 27 10&8.14 &Ap    & 4.69$\pm$0.54&0.114 &0.347 &0.617 &-     &-     &-     &-    &-   &2.03 &4098 &06\\
   &         &        &         &     &      &              &      &      &      &      &      &      &         &    &2.03 &4101 &06 \\
20.&  20880  &03 16 08&-73 32 56&7.95 &Ap    & -            &0.094 &0.208 &1.030 &-     &-     &-     &-     &-   &1.86 &4092 &06\\
21.&  21746 & 03 30 00 & -12 28 39 & 9.41 & ~ K0/K1IV & - &  - &  - &  - &  - &  - &  - &-& -  & 1.07 & 3659 & 05 \\
22.&  21985 & 03 32 25 & -03 18 48 & 8.3 & A1V & 5.63$\pm$0.83 & 0.113 & 0.16 & 0.952 & 2.856 &  - &  - & -& -  & 2.94 & 4104 & 07 \\
 &  &  &  &  &  &  &  &  &  &  &  &  &  &  & 2.27 & 4108 & 07 \\
23.&  22374 &03 36 58&+23 12 40& 6.72&A2p  & 7.65$\pm$0.46& 0.069& 0.178& 1.091& 2.879& 0.022& 0.163& 8397 &1.70 & 2.63 & 4750&08  \\
24.&  22488  &03 32 46&-66 43 46&7.50 &Ap    & 4.39$\pm$0.45&-     &-     &-     &-     &-     &-     &-   &-   &1.93 &4092 &06 \\
25.&  23207  &03 42 44&-18 42 50&7.54 &Ap    & 4.83$\pm$0.71&0.106 &0.259 &0.856 &-     &-     &-     &-    &-   &1.96 &4091 &06 \\
26.&  23393  &03 44 29&-12 03 31&8.30 &F0III & 4.35$\pm$0.91&0.222 &0.164 &0.772 &2.753 & 0.024& 0.133&7271 &1.36&1.96 &4094 &06 \\
27.&  24825  &03 55 16&-38 45 33&6.81 &B9    & 4.23$\pm$0.33&-0.039&0.173 &1.083 &2.835 & 0.035& 0.241&9683 &1.99&1.34 &4077 &06 \\
28.&  25154 & 03 59 48 & -00 01 12 & 9.88 & A5 & 6.74$\pm$1.41 &  - &  - &  - &  - &  - &  - & -& -  & 1.21 & 4397 & 07 \\
29.&  25487 &04 03 54&+28 07 33& 8.08&B8V  & 4.82$\pm$0.99& -    & -    & -    & -    & -    & -    & -   &-& 0.98 & 4459&07     \\
30.&  25999 &04 08 18&+32 27 36& 7.51&Ap   & 6.11$\pm$0.85& -    & -    & -    & -    & -    & -    & -   &-& 1.15 & 4815&09      \\
&  &   &&&& & & & & & && & & 1.52 & 4869&09 \\
&  &   &&&& & & & & & && & &1.43 & 4870&09  \\
31.&  27463  &04 16 21&-60 56 54&6.36 &Ap    & 7.92$\pm$0.42&0.022 &0.224 &0.890 &2.874 &-0.023&-0.028&9214&1.61&1.78 &4091 &06 \\
32.&  28430  &04 27 22&-40 11 50&8.20 &Ap    & 2.52$\pm$0.61&-     &-     &-     &-     &-     &-     &-   &-   &1.89 &4094 &06 \\
33.&  29578  &04 36 31&-54 37 16&8.51 &Ap    & 3.74$\pm$0.61&-     &-     &-     &-     &-     &-     &-    &-   &1.93 &4095 &06 \\
34.&  31225 & 04 53 12 & -20 46 19 & 7.02 & Ap & 5.32$\pm$0.68 & 0.093 & 0.19 & 1.079 &  - &  - &  - & -& -  & 2.33 & 4088 & 06 \\
 &  &  &  &  &  &  &  &  &  &  &  &  &  &  & 1.91 & 4089 & 06 \\
 &  &  &  &  &  &  &  &  &  &  &  &  &  &  & 1.38 & 4090 & 06 \\
35.&  34060  &05 12 03&-49 03 37&7.82 &B9Vp  & 2.74$\pm$0.48&-     &-     &-     &-     &-     &-     &-    &-   &1.91 &4095 &06 \\
36.&  34162 &05 15 31&+05 45 35& 8.68&F0   & 4.79$\pm$1.13& 0.148& 0.186& 0.935& 2.834& 0.027& 0.153& 7982 &1.15 &1.34 & 4400&07   \\
37.&  34205 & 05 15 06 & -15 06 01 & 9.32 & ~ Ap & - & 0.135 & 0.215 & 0.962 & 2.911 &  - &  - &-& -  & 1.60 & 2288 & 02 \\
 &  &  &  &  &  &  &  &  &  &  &  &  &  &  & 1.59 & 2289 & 02 \\
 &  &  &  &  &  &  &  &  &  &  &  &  &  &  & 1.92 & 2296 & 02 \\
 &  &  &  &  &  &  &  &  &  &  &  &  &  &  & 2.14 & 2683 & 03 \\
 &  &  &  &  &  &  &  &  &  &  &  &  &  &  & 2.24 & 2686 & 03 \\
 &  &  &  &  &  &  &  &  &  &  &  &  &  &  & 2.21 & 2693 & 03 \\
38.&  35450 & 05 28 24 & +58 40 29 & 8.16 & A3 & 7.42$\pm$0.87 &  - &  - &  - &  - &  - &  - &-& -  & 1.48 & 4397 & 07 \\
39.&  36955 &05 35 04&-01 24 06&9.58&A2   & -            & 0.057& 0.198& 0.848& 2.866& 0.005&-0.054& 8270  &- & 1.08 & 4427&07    \\
40.&  37308  &05 36 53&-17 00 59&8.71 &A     & -            &-     &-     &-     &-     &-     &-     &-     &-   &1.78 &4096 &06 \\
41.&  38719  &05 44 20&-56 54 58&7.50 &Ap    & 4.19$\pm$0.45&0.011 &0.206 &1.038 &-     &-     &-     &-    &-   &2.19 &4095 &06 \\
42.&  38817 &05 50 37&+44 00 41& 7.56&A2   & 7.27$\pm$0.76& 0.066& 0.217& 0.942& 2.860&-0.012& 0.052& 8209 &1.50 &1.51 & 4071&06 \\
43.&  39082 &05 50 24&+04 57 24& 7.42&B9   & 6.63$\pm$0.53&-0.027& 0.220& 0.887& 2.873&-0.019&-0.029&10451 &2.42 &1.16 & 4428&07  \\
44.&  39575  &05 52 24&-26 17 28&7.83 &A0    & 4.08$\pm$0.70&-0.074&0.267 &0.905 &-     &-     &-     &-   &-   &2.15 &4098 &06 \\
45.&  40277  &05 51 26&-70 28 46&8.33 &Ap    & 4.45$\pm$0.60&0.041 &0.239 &0.901 &-     &-     &-     &-    &-   &1.99 &4102 &07 \\
46.&  40886  &06 00 28&-27 53 18&8.21 &A0    & 0.83$\pm$0.74&-     &-     &-     &-     &-     &-     &-    &-   &2.06 &4096 &06 \\
47.&  41089  &06 00 51&-42 52 14&6.57 &B9IIIp& 4.25$\pm$0.29&-     &-     &-     &-     &-     &-     &-    &-   &3.03 &4092 &06 \\
48.&  41511  &06 04 59&-16 29 04&4.97 &A1V   & 3.59$\pm$0.31&0.186 &0.030 &1.323 &2.775 & 0.164& 0.593&9377 &3.18&1.99 &4103 &07 \\
49.&  41786 &06 08 02&+21 17 44& 7.29&F0   & 9.70$\pm$1.09& 0.193& 0.275& 0.690& 2.782&-0.078& 0.000& 7557 &2.30 &1.11 & 4101&06 \\
50.&  42326  &06 09 17&-17 17 30&7.70 &Ap    & 6.66$\pm$0.66&0.008 &0.231 &0.922 &-     &-     &-     &-   &-   &2.48 &4091 &06 \\
51.&  43901  &06 16 14&-47 49 46&8.20 &Ap    & 1.44$\pm$0.46&0.132 &0.228 &0.946 &2.860 &-0.023& 0.056&8208 &2.35&1.99 &4097 &06 \\
52.&  44195 & 06 20 42 & +05 16 42 & 7.54 & F0 & 11.22$\pm$0.75 & 0.179 & 0.188 & 0.705 & 2.753 &  - &  - &-& -  & 1.96 & 2285 & 02 \\
53.&  44903 & 06 25 20 & +23 03 24 & 8.36 & ~ A5 & - & 0.069 & 0.204 & 0.979 & 2.867 &  - &  - &-& -  & 1.17 & 2285 & 02 \\
54.&  45297 & 06 26 42 & +03 52 18 & 9.23 & ~ B9 & - &  - &  - &  - &  - &  - &  - &-& -  & 2.19 & 4165 & 07 \\
55.&  45698  &06 27 11&-37 06 07&8.15 &A2    & 5.60$\pm$0.56&0.069 &0.244 &0.846 &      &-     &-     &-    &-   &1.99 &4101 &06 \\
56.&  47311 & 06 40 01 & +42 33 55 & 8.71 & F0 & 3.76 & 0.217 & 0.204 & 0.756 & 2.746 &  - &  - &-& - & 1.78 & 1943 & 01 \\
57.&  48953 & 06 46 49 & +16 46 20 & 6.8 & F5 & 10.39 & 0.247 & 0.308 & 0.623 & 2.752 &  - &  - & -& - & 1.10 & 2210 & 01 \\
 &  &  &  &  &  &  &  &  &  &  &  &  &   &  & 0.91 & 2305 & 02 \\
58.&  51496 &07 00 57&+56 51 13& 9.83&F5   & -            & -    & -    & -    & -    & -    & -    & -    &-   &2.01 & 4106&07    \\
59.&  51684 & 06 56 29 & -40 59 25 & 7.94 & Ap & 3.58$\pm$0.60 & 0.154 & 0.248 & 0.768 & 2.832 &  - &  - & -& -  & 2.10 & 4088 & 06 \\
60.&  55719  &07 12 16&-40 29 56&5.31 &A3spe & 7.93$\pm$0.38&0.012 &0.217 &1.030 &2.880 &-0.017& 0.100&9101 &2.28&2.09 &4102 &07 \\
61.&  56148 &07 19 48&+61 35 29& 9.00&F0   & -            & 0.204& 0.178& 0.628& 2.722& 0.000& 0.044& 7046  &-  &1.05 & 4104&07    \\
62.&  56350  &07 13 40&-53 40 04&6.69 &Ap    & 6.61$\pm$0.26&-     &-     &-     &2.799 &-     &-     &-   &-   &2.45 &4096 &06 \\
63.&  61763  &07 38 47&-44 49 48&7.94 &Apsh  & 2.54$\pm$0.41&-     &-     &-     &-     &-     &-     &-    &-   &2.35 &4097 &06\\
64.&  66195  &07 56 47&-70 42 59&8.65 &Ap    & 5.17$\pm$0.76&0.043 &0.227 &0.886 &-     &-     &-     &-    &-   &2.03 &4103 &07 \\
65.&  70338 & 08 21 53 & +13 37 26 & 7.32 & A2 & 5.51$\pm$0.76 & 0.173 & 0.271 & 0.805 & 2.814 &  - &  - & -& 1.45  & 0.93 & 2338 & 02 \\
66.&  72611  &08 32 17&-41 49 56&7.01 &Ap    & 5.61$\pm$0.42&-0.062&0.192 &0.748 &-     &-     &-     &-    &-   &2.27 &4098 &06 \\
   &         &        &         &     &      &              &      &      &      &      &      &      &     &    &0.96 &4101 &06 \\
67.&  72634  &08 29 43&-67 08 23&7.27 &Ap    & 3.35$\pm$0.45&-0.011&0.185 &1.025 &-     &-     &-     &-    &-   &2.01 &4103 &07 \\
68.&  72943 &08 36 08&+15 18 49& 6.32&F0IV &12.86$\pm$0.45& 0.211& 0.186& 0.732& 2.720&-0.009& 0.152& 6991 &1.16 &1.13 & 4099&06 \\
69.&  73095 & 08 37 35 & +31 50 31 & 8.85 & ~ A3 & - & 0.19 & 0.195 & 0.702 & 2.746 &  - &  - &-& -  & 0.96 & 2239 & 01 \\
70.&  73345 &08 38 38&+19 59 23& 8.14&F0V  & -            & 0.121& 0.210& 0.883& 2.812&-0.004& 0.140& 7780  &-  &1.19 & 4429&07    \\
71.&  73574 &08 39 43&+20 05 11& 7.75&A5V  & -            & 0.127& 0.207& 0.871& 2.799&-0.004& 0.093& 7659 &-  &2.47 & 4164&07   \\
&  &   &&&& & & & & & && & &1.61 & 4166 &07 \\
72.&  74067  &08 40 19&-40 15 50&5.20 &B9V   &11.68$\pm$0.50&-0.050&0.220 &0.898 &2.846 &-0.013& 0.036&10412&-&2.38 &4102 &07 \\
73.&  75445 & 08 48 42 & -39 14 01 & 7.12 & A3 & 9.23$\pm$0.45 & 0.159 & 0.218 & 0.729 &  - &  - &  - & -& -  & 2.80 & 2288 & 02 \\
 &  &  &  &  &  &  &  &  &  &  &  &  &  &  & 1.45 & 2289 & 02 \\
 &  &  &  &  &  &  &  &  &  &  &  &  &  &  & 1.97 & 2296 & 02 \\
 &  &  &  &  &  &  &  &  &  &  &  &  &  &  & 1.99 & 2704 & 03 \\
 &  &  &  &  &  &  &  &  &  &  &  &  &  &  & 2.04 & 2709 & 03 \\
 &  &  &  &  &  &  &  &  &  &  &  &  &  &  & 2.16 & 2710 & 03 \\
74.&  76444 &08 57 07&+29 12 57& 9.11&F0   & 3.11$\pm$1.02& 0.176& 0.191& 0.752& 2.745&-0.006& 0.128& 7204 &1.27&1.44 & 4104&07   \\
75.&  78388 &09 09 52&+49 49 56& 7.61&F0III& 9.25$\pm$0.66& 0.231& 0.172& 0.710& 2.709& 0.002& 0.153& 6900 &0.92  & 2.28 & 4103&07\\
76.&  82417 & 09 30 22 & -46 48 48 & 9.24 & ~ Ap & - &  - &  - &  - &  - &  - &  - &-& -  & 5.14 & 3483 & 05 \\
77.&  86170 &09 56 45&-02 17 20& 8.42&A2   & 3.20$\pm$0.98& 0.074& 0.226& 0.929& -    & -    & -    & -    &- & 1.39 & 4428&07    \\
&  &   &&&& & & & & & && & &1.77 & 4431&07 \\
&  &   &&&& & & & & & && & &1.55 & 4459&07 \\
78.&  88385 & 10 09 49 & -56 44 53 & 8.09 & Ap & 4.85$\pm$0.55 & 0.006 & 0.23 & 0.863 & 2.822 &  - &  - &-& -  & 1.94 & 4174 & 07 \\
79.&  88701 & 10 13 00 & -37 30 12 & 9.27 & B9 & 2.36$\pm$0.99 &  - &  - &  - &  - &  - &  - & -& -  & 1.34 & 4175 & 07 \\
80.&  100809 & 11 36 14 & +14 41 51 & 8.25 & Am & 7.12$\pm$0.85 & 0.1 & 0.269 & 0.856 & 2.845 &  - &  - & -& -  & 1.41 & 2009 & 01 \\
81.&  104044 & 11 58 53 & -43 22 55 & 9.57 & ~ Ap & - &  - &  - &  - &  - &  - &  - & -& -  & 0.91 & 3482 &05  \\
82.&  106374 & 12 14 18 & -33 46 44 & 7.37 & A2 & 5.63$\pm$0.54 &  - &  - &  - &  - &  - &  - &-& -  & 0.97 & 3482 & 05 \\
83.& 117044 &13 27 30&+13 54 49& 8.19&F0II & 5.25$\pm$0.76& 0.209& 0.211& 0.674& 2.749&-0.025& 0.043& 7267 &1.19&1.32 & 4104&07  \\
84.&  117290 & 13 30 13 & -49 07 58 & 9.25 & ~ Ap & - &  - &  - &  - &  - &  - &  - &-& -  & 1.97 & 4174 & 07 \\
 &  &  &  &  &  &  &  &  &  &  &  &  &   &  & 3.22 & 4175 & 07 \\
 &  &  &  &  &  &  &  &  &  &  &  &  &   &  & 1.41 & 4178 & 07 \\
 &  &  &  &  &  &  &  &  &  &  &  &  &   &  & 1.50 & 4179 & 07 \\
 &  &  &  &  &  &  &  &  &  &  &  &  &   &  & 2.48 & 4180 & 07 \\
 &  &  &  &  &  &  &  &  &  &  &  &  &   &  & 3.48 & 4181 & 07 \\
85.&  127608 & 14 33 47 & -46 45 33 & 8.56 & ~ Ap & - &  - &  - &  - &  - &  - &  - &-& -  & 4.66 & 3515 & 05 \\
86.&  140220 & 15 44 10 & -44 06 50 & 7.97 & ~ Ap & - &  - &  - &  - &  - &  - &  - &-& -  & 2.01 & 3483 & 05 \\
87.&  144897 & 16 09 51 & -41 09 27 & 8.59 & Ap & 5.61$\pm$1.04 &  - &  - &  - &  - &  - &  - & -& - & 2.01 & 2507 & 02 \\
 &  &  &  &  &  &  &  &  &  &  &  &  &   &  & 1.60 & 2511 & 02 \\
 &  &  &  &  &  &  &  &  &  &  &  &  &   &  & 1.71 & 2513 & 02 \\
 &  &  &  &  &  &  &  &  &  &  &  &  &   &  & 1.63 & 2516 & 02 \\
 &  &  &  &  &  &  &  &  &  &  &  &  &   &  & 1.58 & 2520 & 02 \\
88.&  149769 & 16 40 47 & -62 25 54 & 9.75 & ~ Ap & - &  - &  - &  - &  - &  - &  - & -& -  & 3.79 & 2123 & 01 \\
89.&  161423 & 17 52 02 & -71 41 24 & 9.31 & ~ Ap & - &  - &  - &  - &  - &  - &  - & -& - & 3.37& 2127 & 01 \\
 &  &  &  &  &  &  &  &  &  &  &  &  &   &  & 1.24 & 2128 & 01 \\
90.&  162639 & 17 54 41 & -50 26 45 & 9.93 & ~ Ap & - &  - &  - &  - &  - &  - &  - & -& -  & 1.06 & 3482 & 05 \\
91.&  164258 & 18 00 15 & +00 37 46 & 6.37 & A3spe & 7.39$\pm$0.52 & 0.087 & 0.181 & 1.099 & 2.905 &  - &  - & -& -  & 0.53 & 3518 & 05 \\
92.&  168767 & 18 22 30 & -26 54 40 & 8.71 & ~ A0 & - &  - &  - &  - &  - &  - &  - & -& -  & 1.50 & 3482 & 05 \\
93.&  169380 & 18 26 06 & -37 54 42 & 9.83 & ~ A3 & - &  - &  - &  - &  - &  - &  - & -& -  & 2.02 & 2128 & 01 \\
94.& 170397  &18 29 47&-14 34 55&6.02 &Ap    & 9.54$\pm$0.36&-0.029&0.190 &0.925 &2.837 & 0.018& 0.080&10447&3.85&1.98 &4298 &07\\
95.& 172976 &18 41 03&+44 16 16&7.29&F0III& 5.03$\pm$0.48& 0.181& 0.234& 0.827& 2.788&-0.035& 0.126& 7578 &1.62  &2.03 & 4251&07\\
&  &   &&&& & & & & & && & &1.11 & 4374&07\\
96.&  173612 & 18 46 30 & -08 25 58 & 9.08 & ~ A0 & - &  - &  - &  - &  - &  - &  - & -& -  & 2.74 & 3515 & 05 \\
 &  &  &  &  &  &  &  &  &  &  &  &  &  &  &  &  &  \\
97.& 178892 &19 09 55&+14 57 58& 8.94&B9   & 4.53$\pm$1.10& -    & -    & -    & -    & -    & -    & -    &- &0.86 & 4375&07   \\
98.& 183806  &19 33 22&-45 16 18&5.58 &Ap    & 8.22$\pm$0.40&-0.025&0.167 &1.062 &2.849 & 0.039& 0.194&9816 &2.01&2.59 &4295 &07\\
99.& 187474  &19 51 51&-39 52 28&5.32 &Ap    &10.82$\pm$0.88&-0.047&0.203 &0.864 &2.820 & 0.003& 0.044&10706&1.93&3.20 &4294 &07 \\
100.& 188008  &19 54 27&-36 34 32&8.86 &A5    & -            &0.040 &0.248 &0.858 &2.880 &-0.048&-0.072&8797  &-   &1.43 &4296 &07\\
101.& 190401 &20 03 09&+41 28 28& 6.99&Am   & 9.34$\pm$0.35&0.220 & 0.209& 0.728& 2.744&-0.026& 0.059& 7204 &2.13 &1.13 & 4374&07\\
&  &   &&&& & & & & & && & &2.40 & 4375&07\\
&  &   &&&& & & & & & && &&2.01 & 4376&07 \\
102.&  196604 & 20 36 50 & +44 54 40 & 8.12 & A3 & 8.04$\pm$0.70 & 0.222 & 0.218 & 0.633 & 2.737 &  - &  - &-& - & 1.07 & 1832 & 00 \\
103.& 204367  &21 28 41&-25 38 39&7.83 &A0    & 5.07$\pm$0.77&-     &-     &-     &-     &-     &-     &-    &-   &2.09 &4295 &07\\
104.& 205087 &21 32 27&+23 23 40& 6.68&B9sp & 5.87$\pm$0.36&-0.064& 0.189& 0.752& 2.799& 0.014&-0.589&11906 &1.94 &1.35 & 4428&07\\
105.&  208759 & 22 00 54 & -64 57 41 & 9.98 & ~ Ap & - &  - &  - &  - &  - &  - &  - &-& - & 2.02 & 2127 & 01 \\
 &  &  &  &  &  &  &  &  &  &  &  &  &  &  & 1.67 & 2128 & 01 \\
106.& 212385  &22 24 38&-39 07 37&6.84 &A2p   & 7.92$\pm$0.63&0.067 &0.225 &0.946 &-     &-     &-     &-    &-   &2.04 &4294 &07\\
107.&  216018 & 22 49 26 & -11 20 57 & 7.62 & A7 & 8.03$\pm$0.66 & 0.165 & 0.318 & 0.561 &  - &  - &  - & -& -& 1.35 & 2574 & 02 \\
 &  &  &  &  &  &  &  &  &  &  &  &  &   &  & 2.50 & 2585 & 02 \\
 &  &  &  &  &  &  &  &  &  &  &  &  &   &  & 2.69 & 2586 & 02 \\
 &  &  &  &  &  &  &  &  &  &  &  &  &   &  & 2.24 & 2588 & 02 \\
 &  &  &  &  &  &  &  &  &  &  &  &  &   &  & 2.16 & 2589 & 02 \\
 &  &  &  &  &  &  &  &  &  &  &  &  &   &  & 2.17 & 2590 & 02 \\
108.& 290665 &05 35 10&-00 50 13& 9.44&B9   & -            & 0.037& 0.207& 0.829& 2.854&-0.001&-0.049& 9167 &-   &1.34 & 4459&07   \\
&  &   &&&& & & & & & && &&1.94 & 4492&08 \\
\\
\hline
\end{longtable}
\label{null1}
\end{landscape}

\onecolumn
\begin{figure}
\centering
\includegraphics[width=0.95\linewidth]{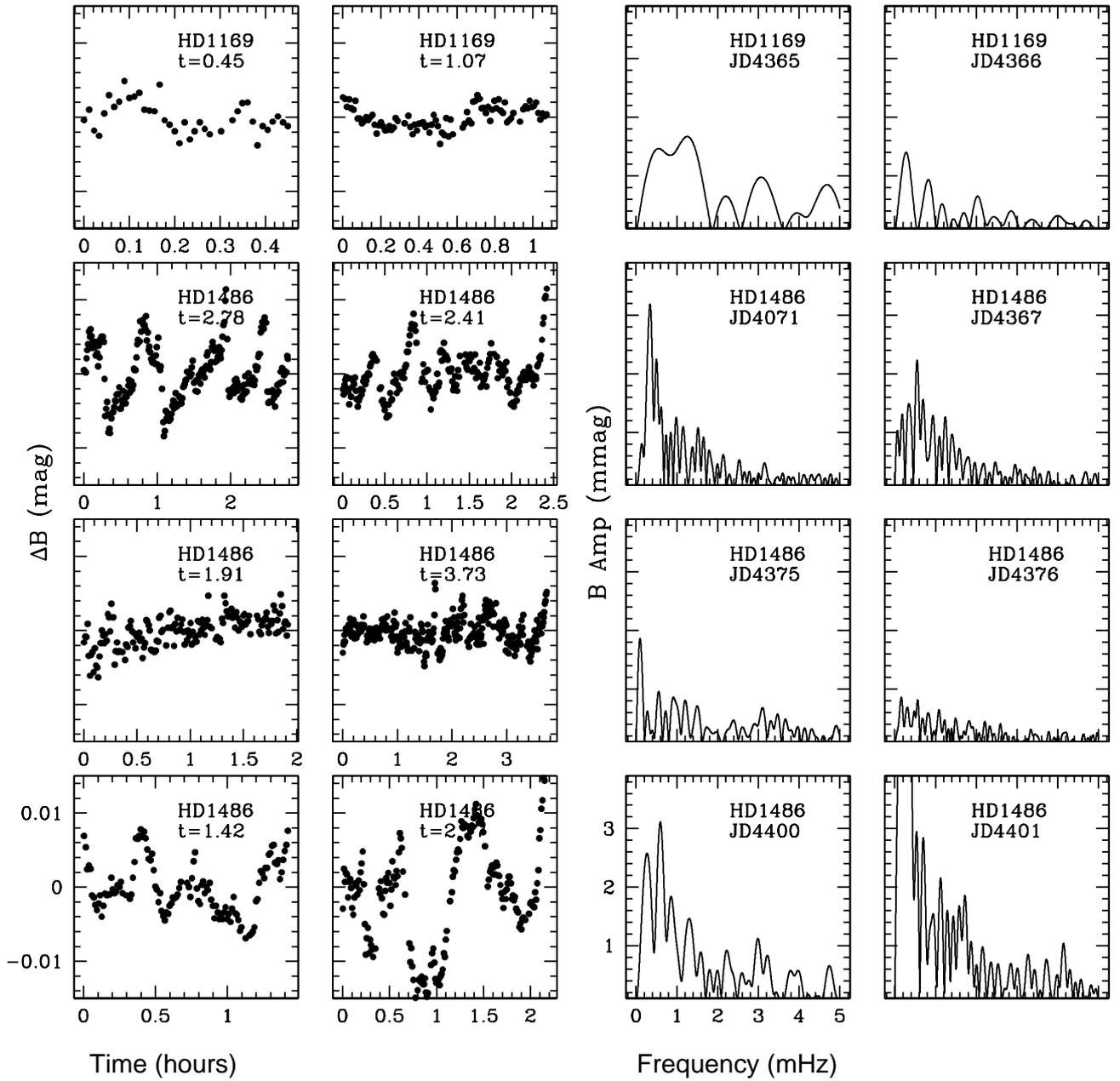}
\caption{The light curves (left columns) of the pulsation candidate stars observed from ARIES/SAAO and their corresponding pre-whitened amplitude spectra (right columns). The light curves have been binned to 40-sec integrations.  }
\label{null1}
\end{figure}
 
\addtocounter{figure}{-1}  
    \begin{figure}
   \centering
   \includegraphics[width=0.95\linewidth]{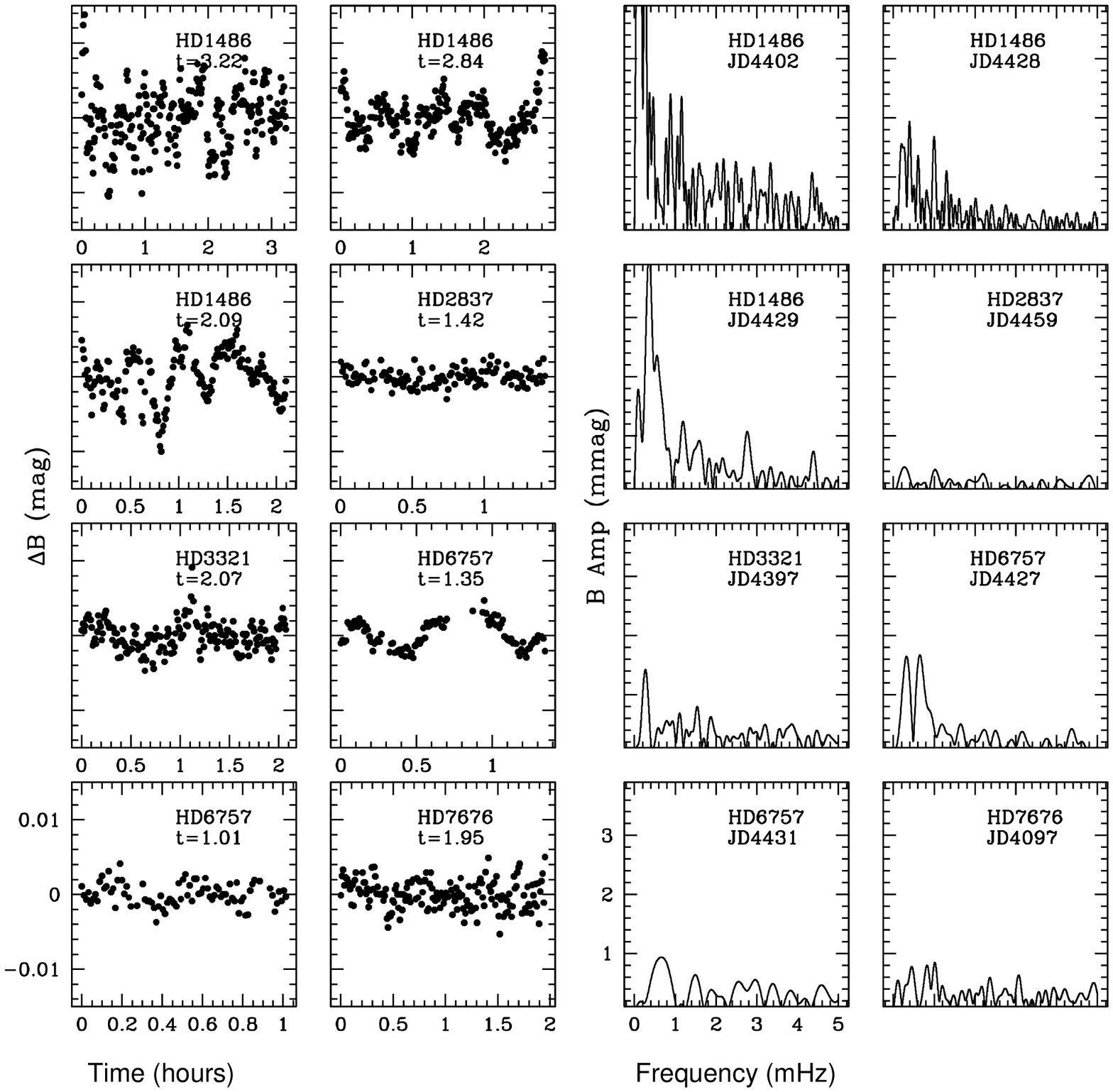}
   \caption{Continued.}
   \end{figure} 
   
  \addtocounter{figure}{-1}   
   \begin{figure}
   \centering
   \includegraphics[width=0.95\linewidth]{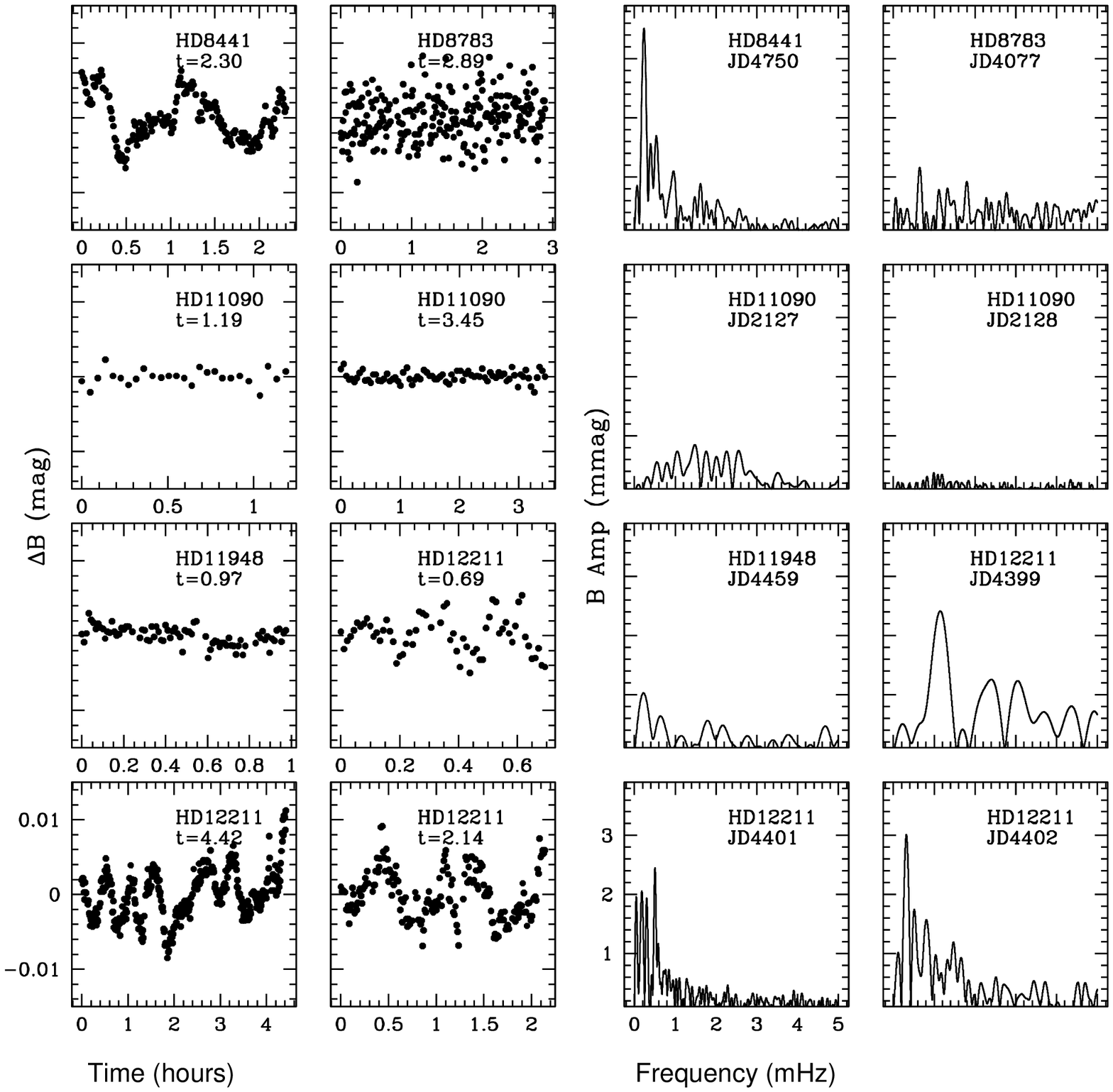}
    \caption{Continued.}
   \end{figure}
   
     \addtocounter{figure}{-1}   
    \begin{figure}
   \centering
   \includegraphics[width=0.95\linewidth]{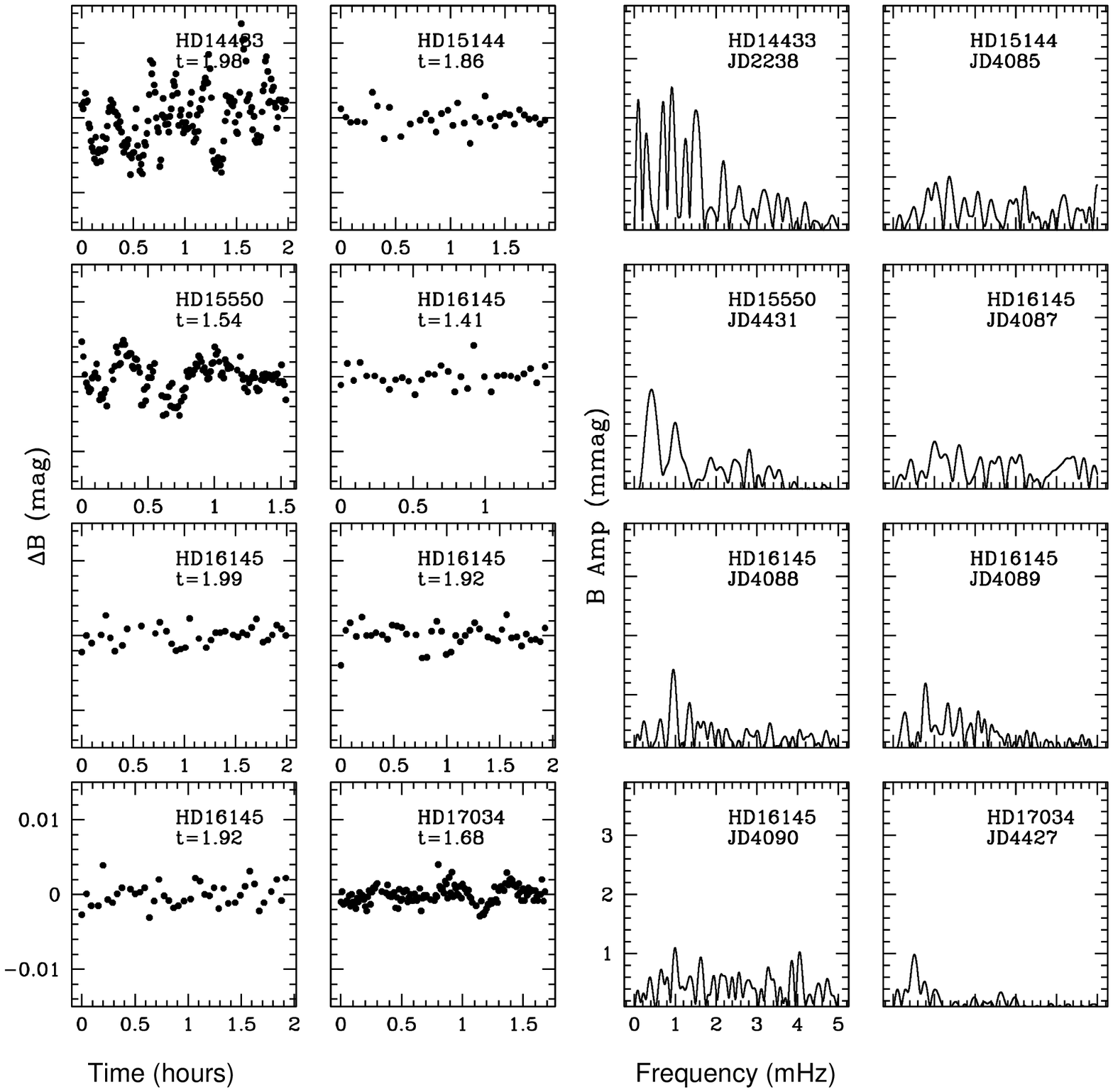}
    \caption{Continued.}
   \end{figure}
   
   \addtocounter{figure}{-1}
  \begin{figure}
   \centering
    \includegraphics[width=0.95\linewidth]{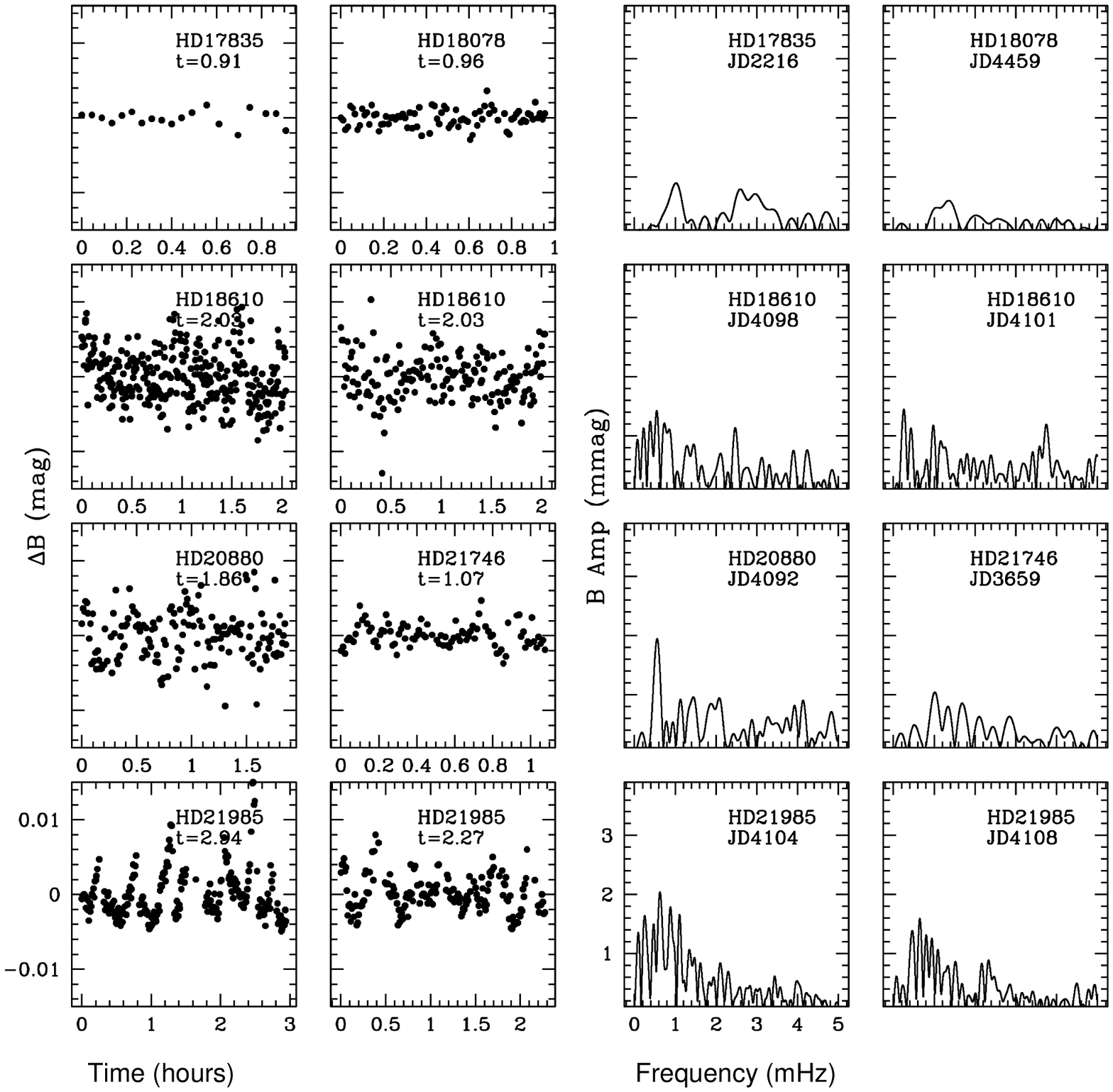}
    \caption{Continued.}
\end{figure}
      
      \addtocounter{figure}{-1}
    \begin{figure}
   \centering
   \includegraphics[width=0.95\linewidth]{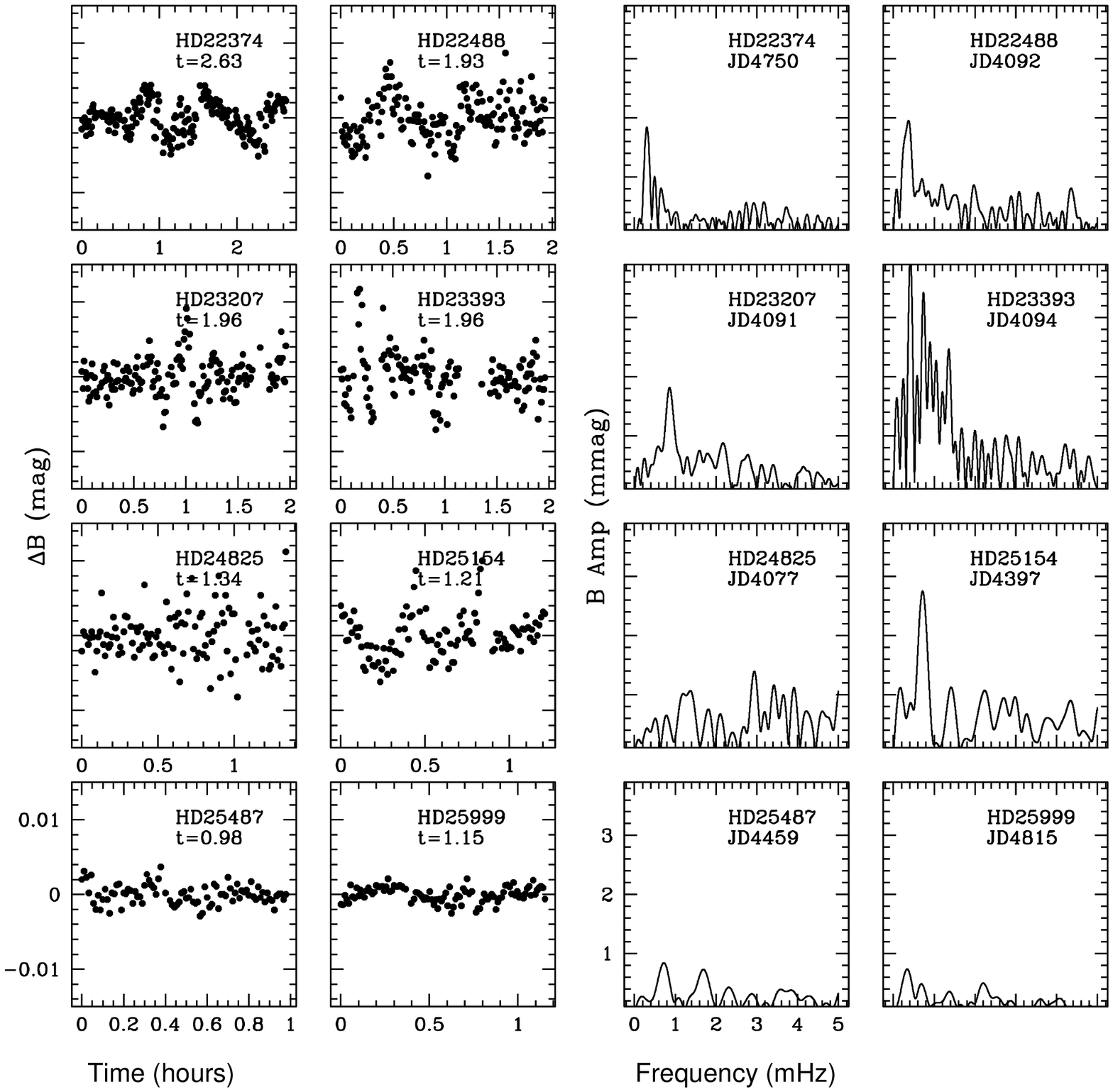}
   \caption{Continued.}
   \end{figure} 
   
    \addtocounter{figure}{-1}
   \begin{figure}
   \centering
   \includegraphics[width=0.95\linewidth]{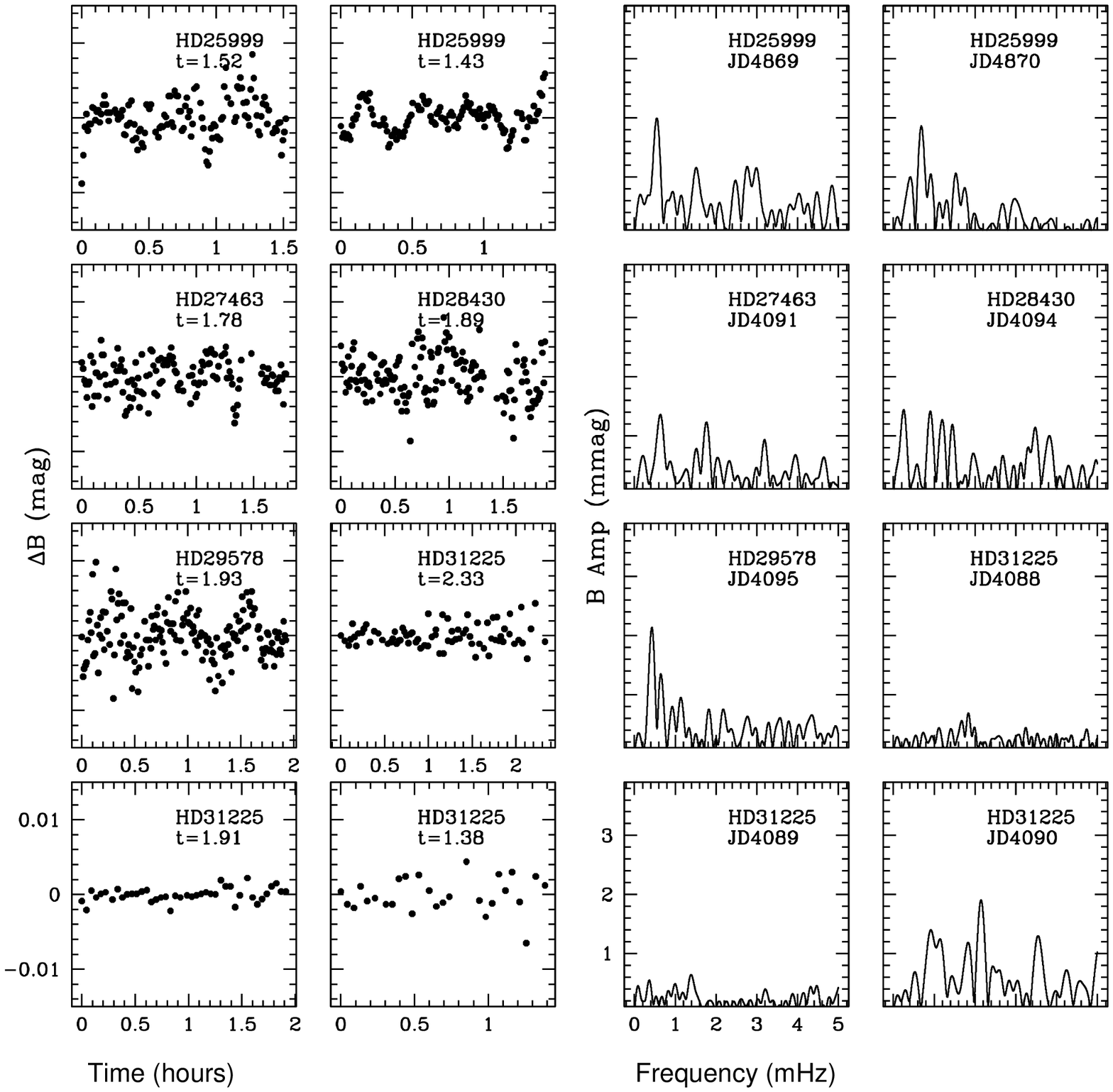}
    \caption{Continued.}
   \end{figure}
   
   \addtocounter{figure}{-1}
    \begin{figure}
   \centering
   \includegraphics[width=0.95\linewidth]{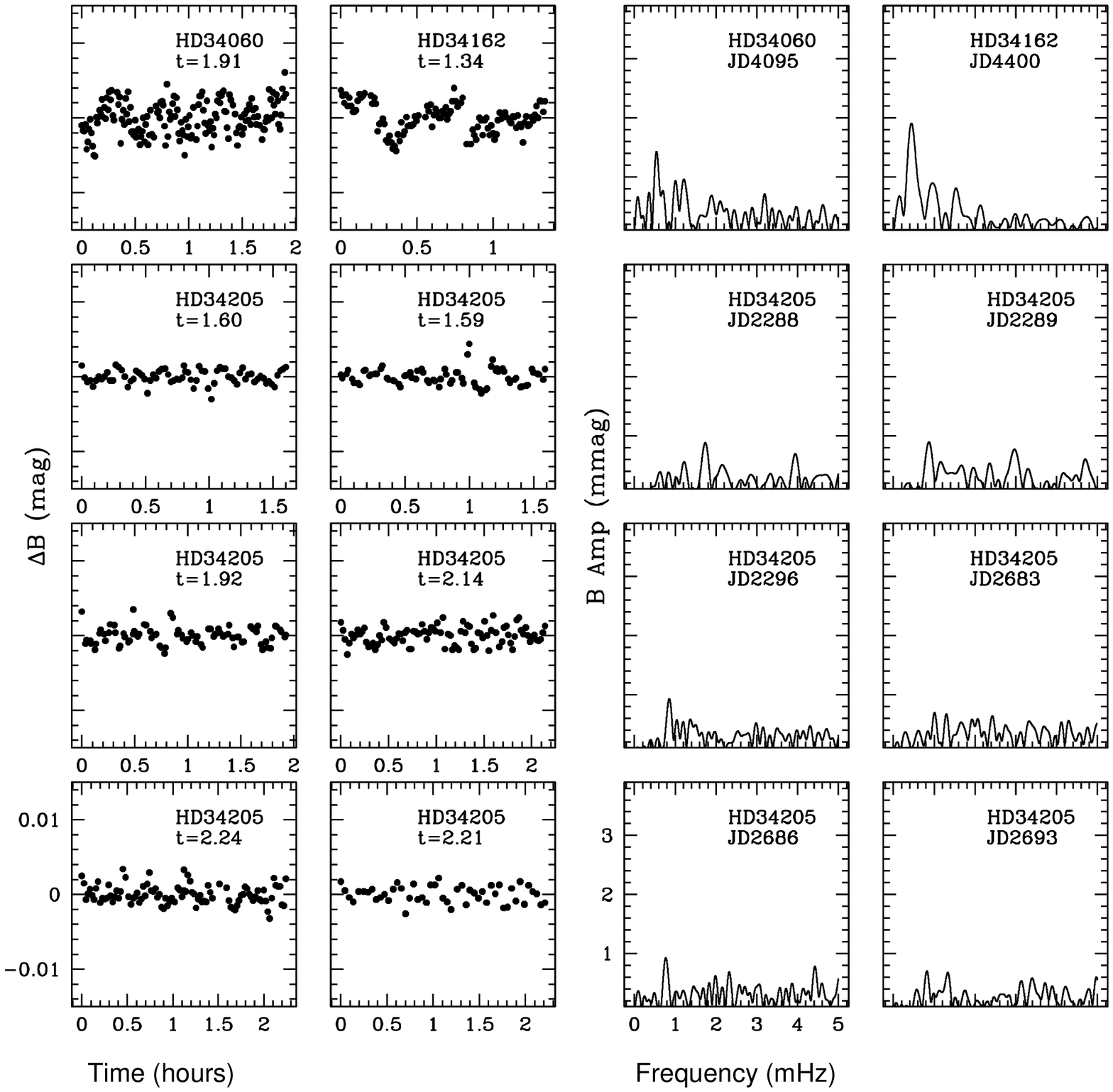}
    \caption{Continued.}
   \end{figure}
      
      \addtocounter{figure}{-1}
  \begin{figure}
   \centering
    \includegraphics[width=0.95\linewidth]{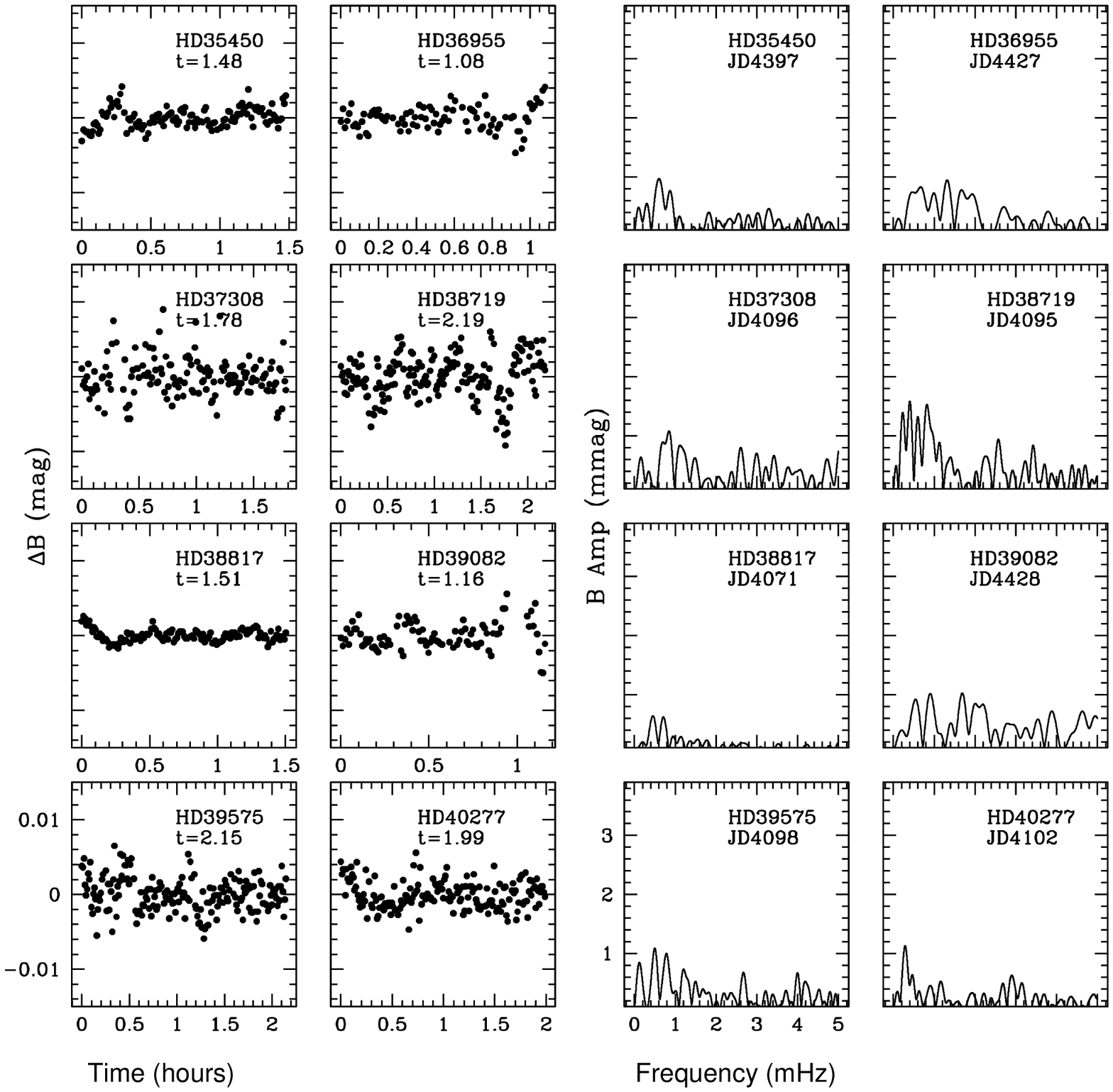}
    \caption{Continued.}
\end{figure}   
   
      \addtocounter{figure}{-1}  
    \begin{figure}
   \centering
   \includegraphics[width=0.95\linewidth]{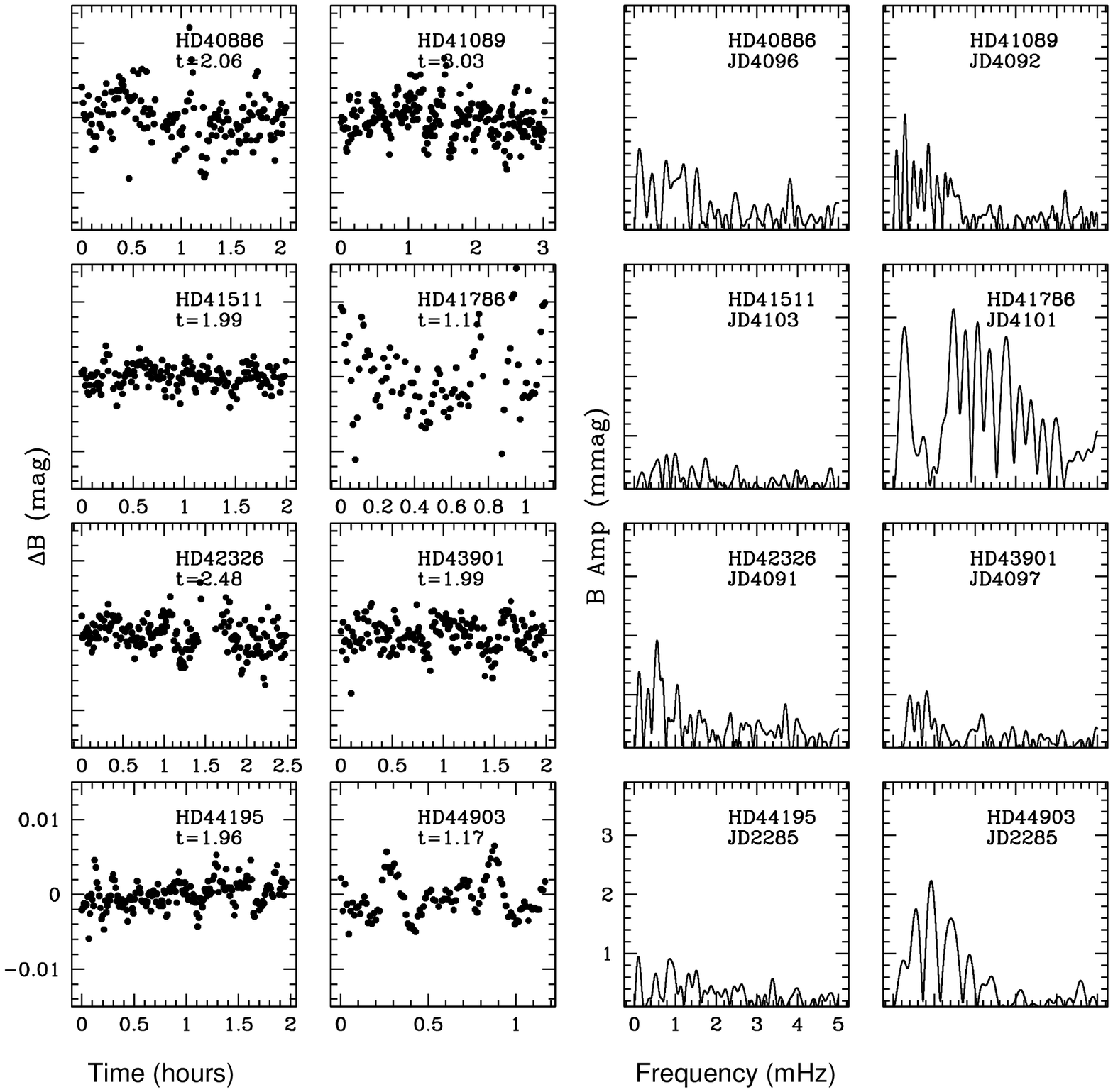}
   \caption{Continued.}
   \end{figure} 
   
  \addtocounter{figure}{-1}   
   \begin{figure}
   \centering
   \includegraphics[width=0.95\linewidth]{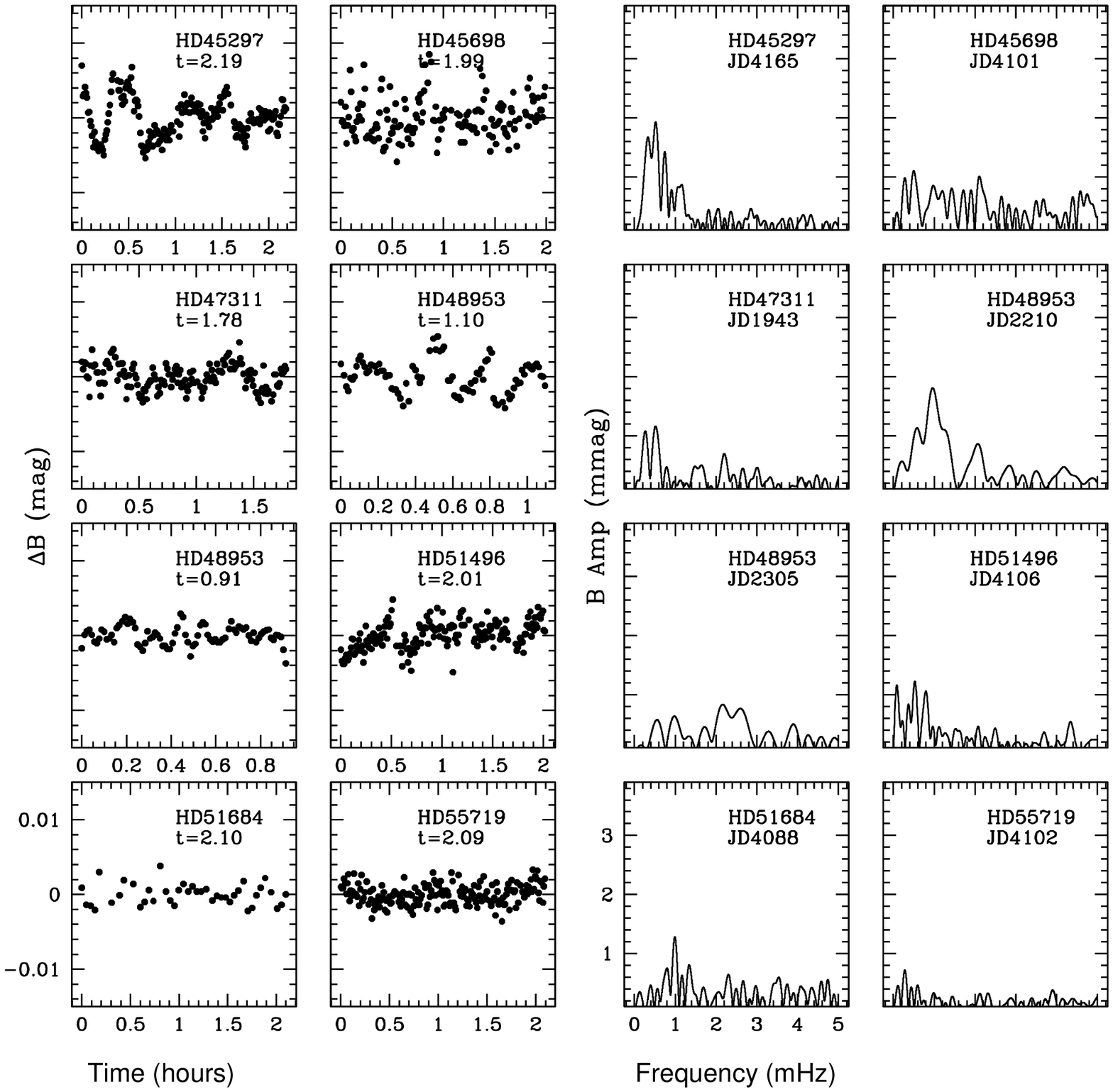}
    \caption{Continued.}
   \end{figure}
   
     \addtocounter{figure}{-1}   
    \begin{figure}
   \centering
   \includegraphics[width=0.95\linewidth]{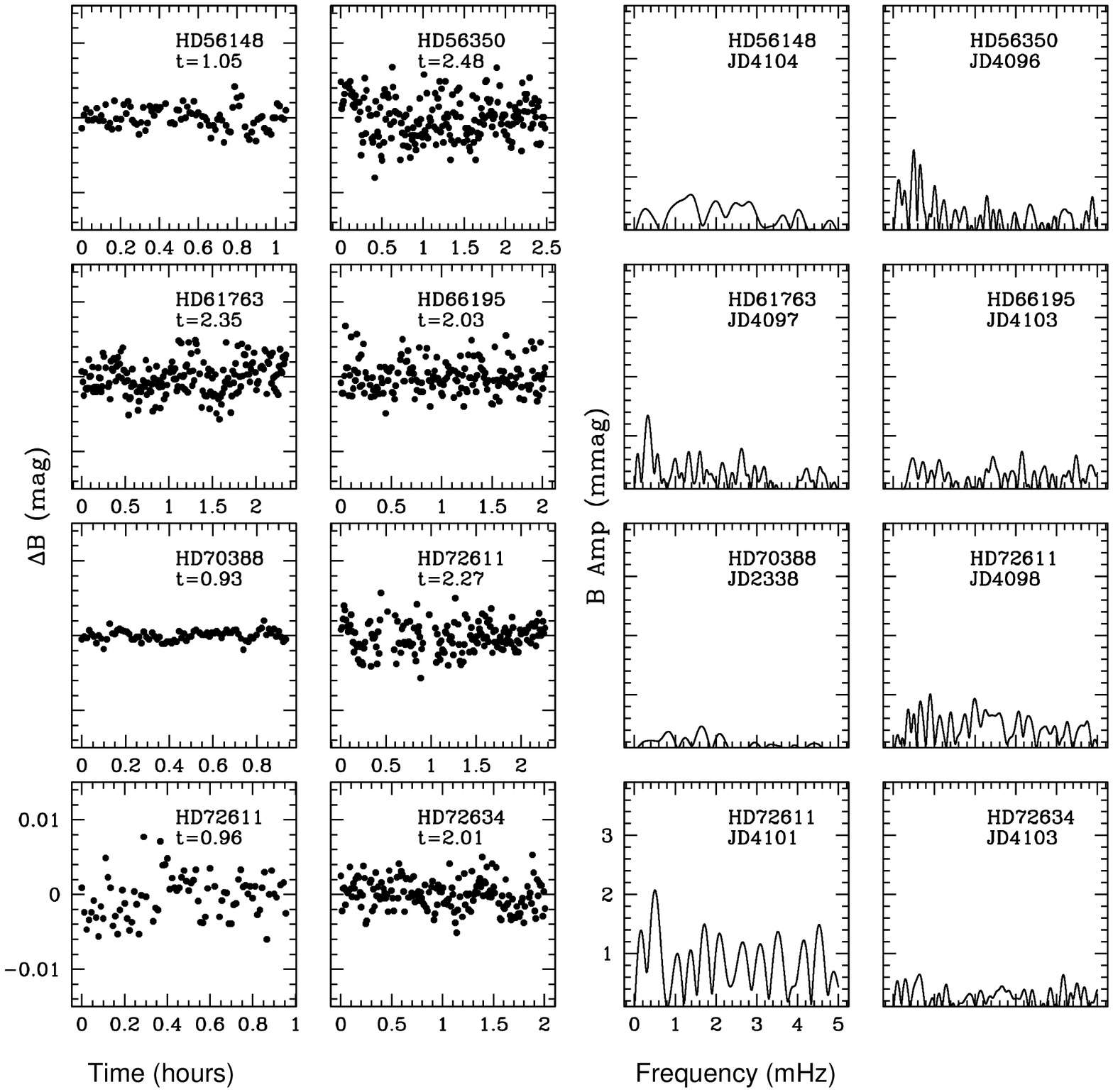}
    \caption{Continued.}
   \end{figure}
   
   \addtocounter{figure}{-1}
  \begin{figure}
   \centering
    \includegraphics[width=0.95\linewidth]{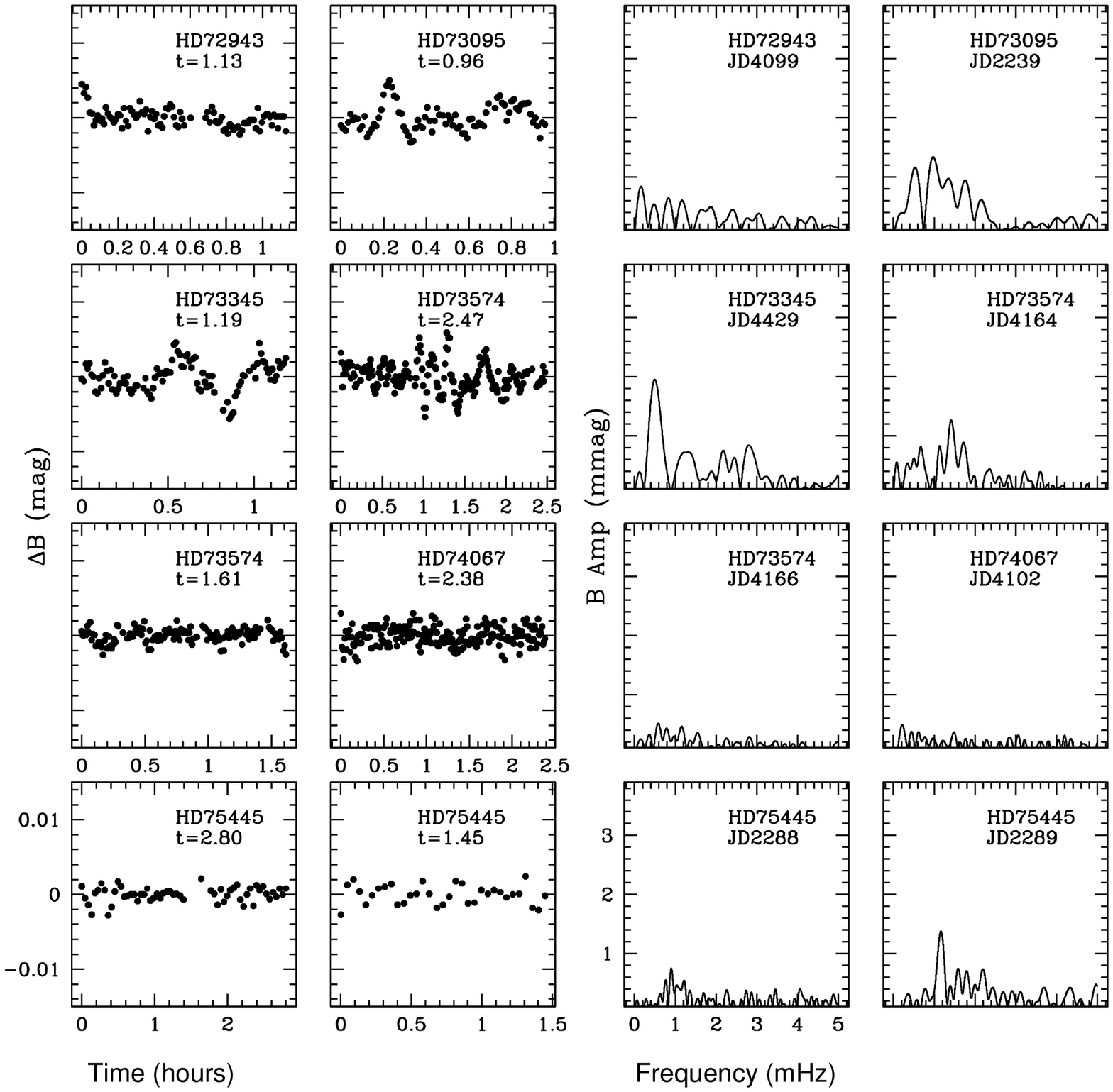}
    \caption{Continued.}
   \end{figure}
   
   \addtocounter{figure}{-1}  
    \begin{figure}
   \centering
   \includegraphics[width=0.95\linewidth]{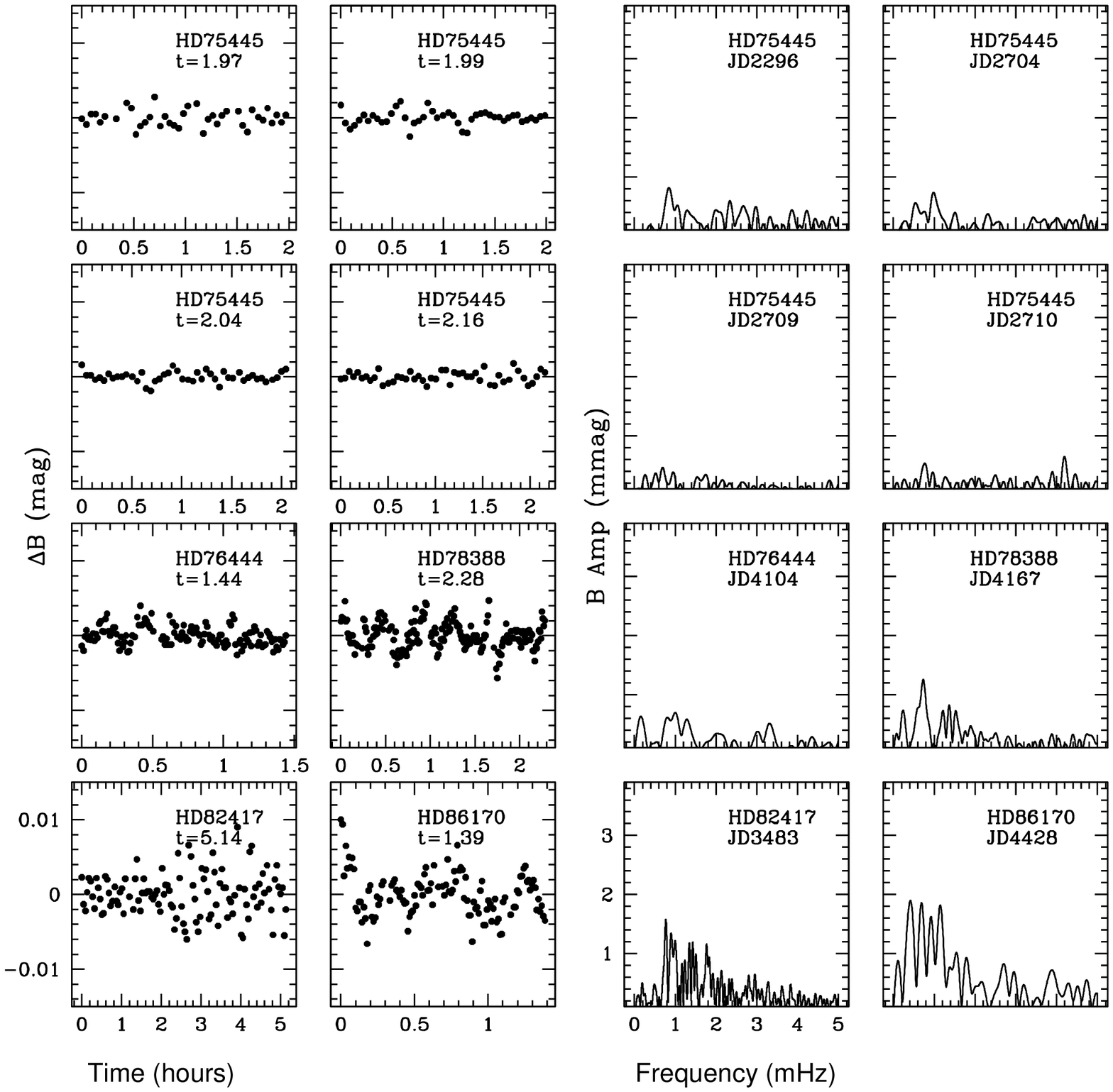}
   \caption{Continued.}
   \end{figure} 
   
  \addtocounter{figure}{-1}   
   \begin{figure}
   \centering
   \includegraphics[width=0.95\linewidth]{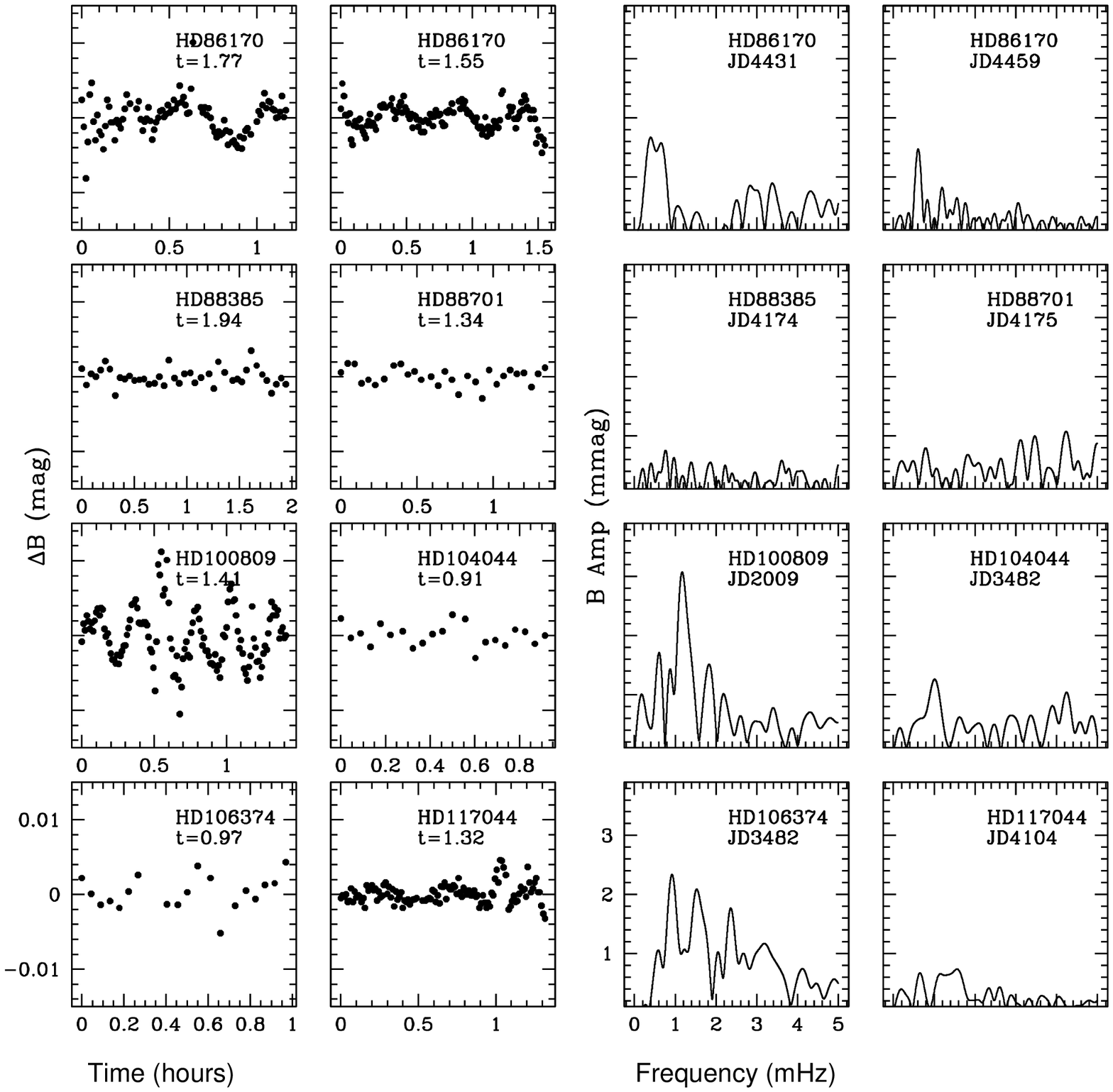}
    \caption{Continued.}
   \end{figure}
   
     \addtocounter{figure}{-1}   
    \begin{figure}
   \centering
   \includegraphics[width=0.95\linewidth]{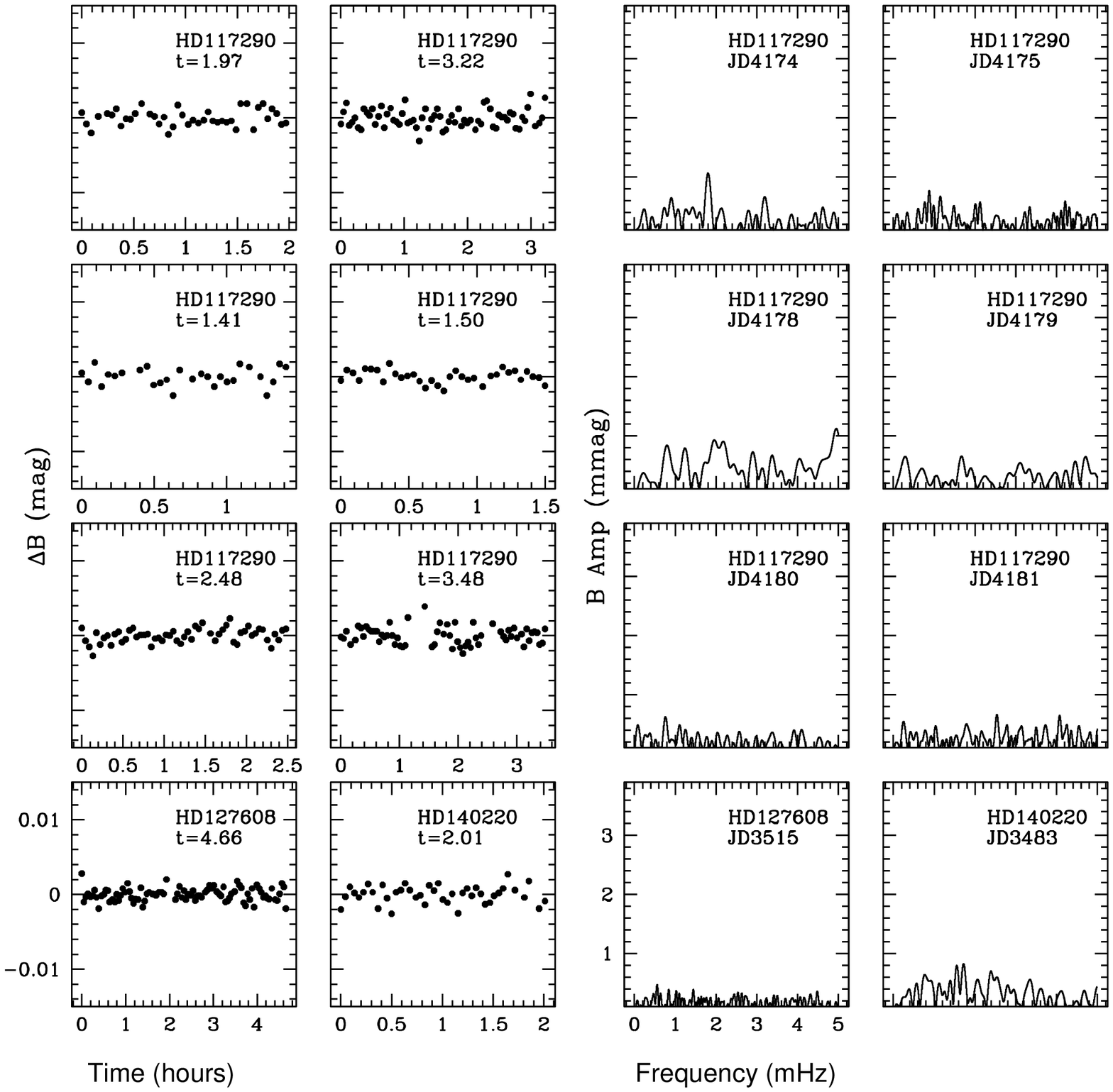}
    \caption{Continued.}
   \end{figure}
   
   \addtocounter{figure}{-1}
  \begin{figure}
   \centering
    \includegraphics[width=0.95\linewidth]{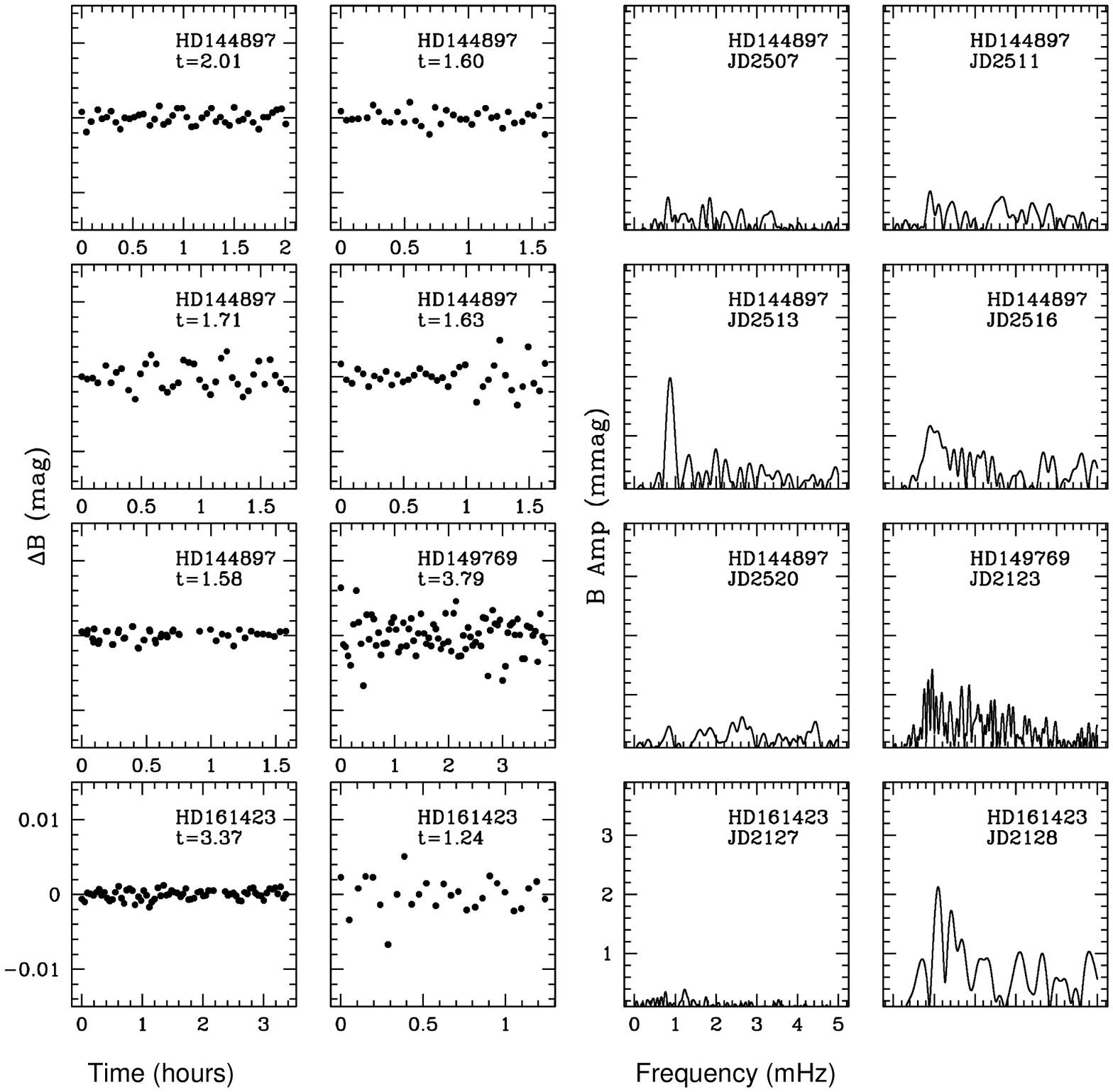}
    \caption{Continued.}
   \end{figure}

   \addtocounter{figure}{-1}  
    \begin{figure}
   \centering
   \includegraphics[width=0.95\linewidth]{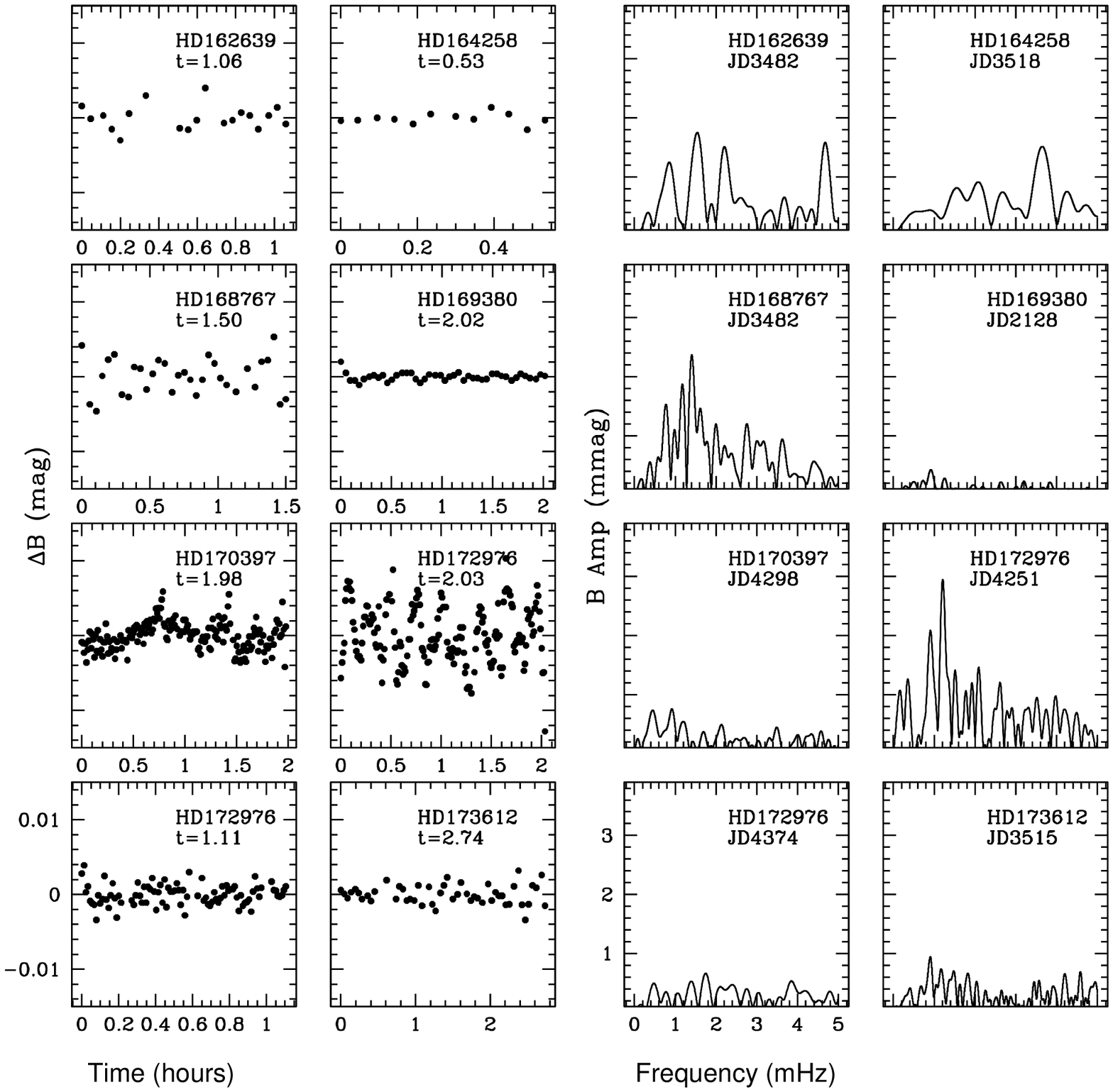}
   \caption{Continued.}
\end{figure} 
\twocolumn

\onecolumn
\addtocounter{figure}{-1}   
   \begin{figure}
   \centering
   \includegraphics[width=0.95\linewidth]{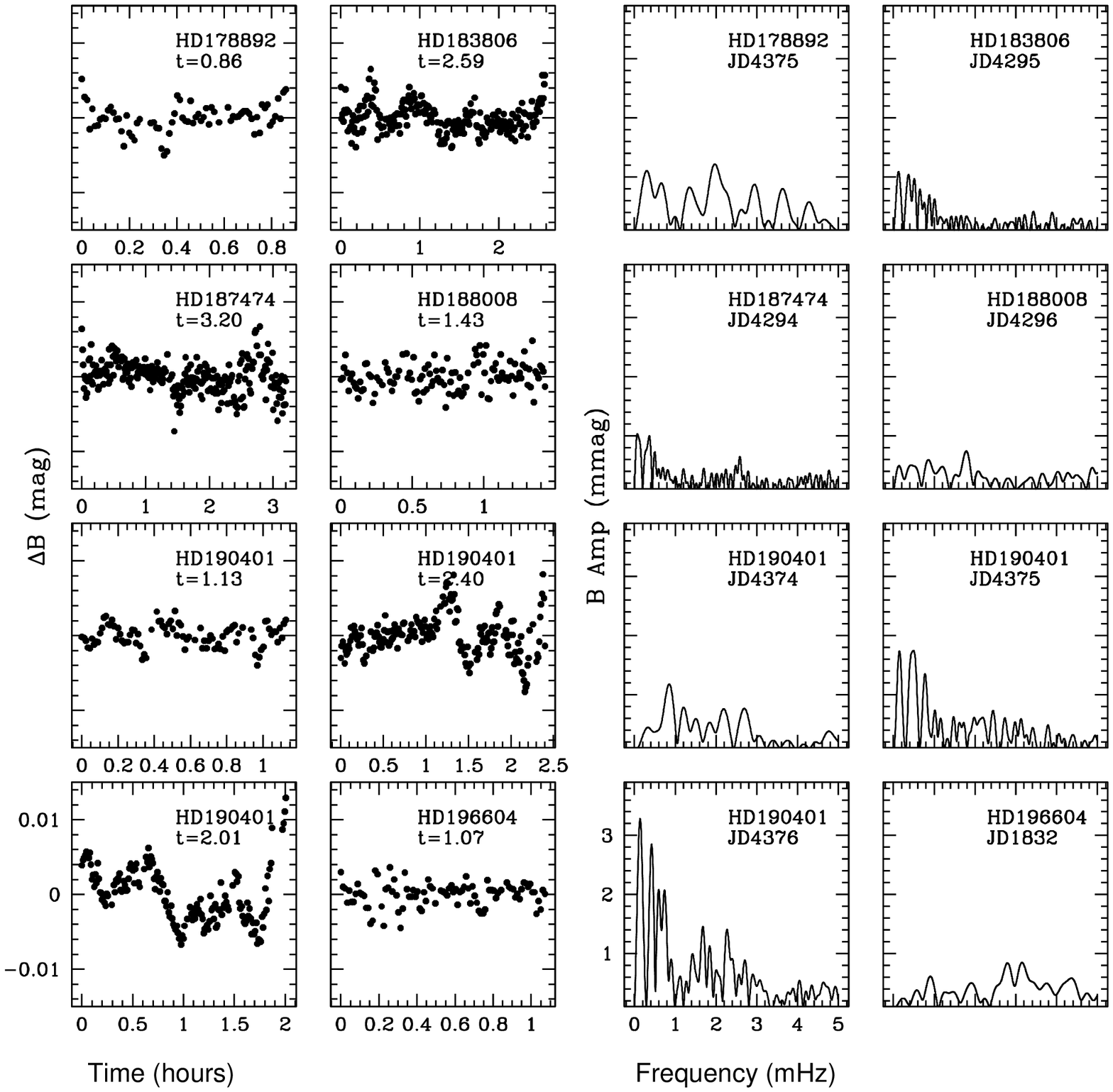}
    \caption{Continued.}
   \end{figure}
   
     \addtocounter{figure}{-1}   
    \begin{figure}
   \centering
   \includegraphics[width=0.95\linewidth]{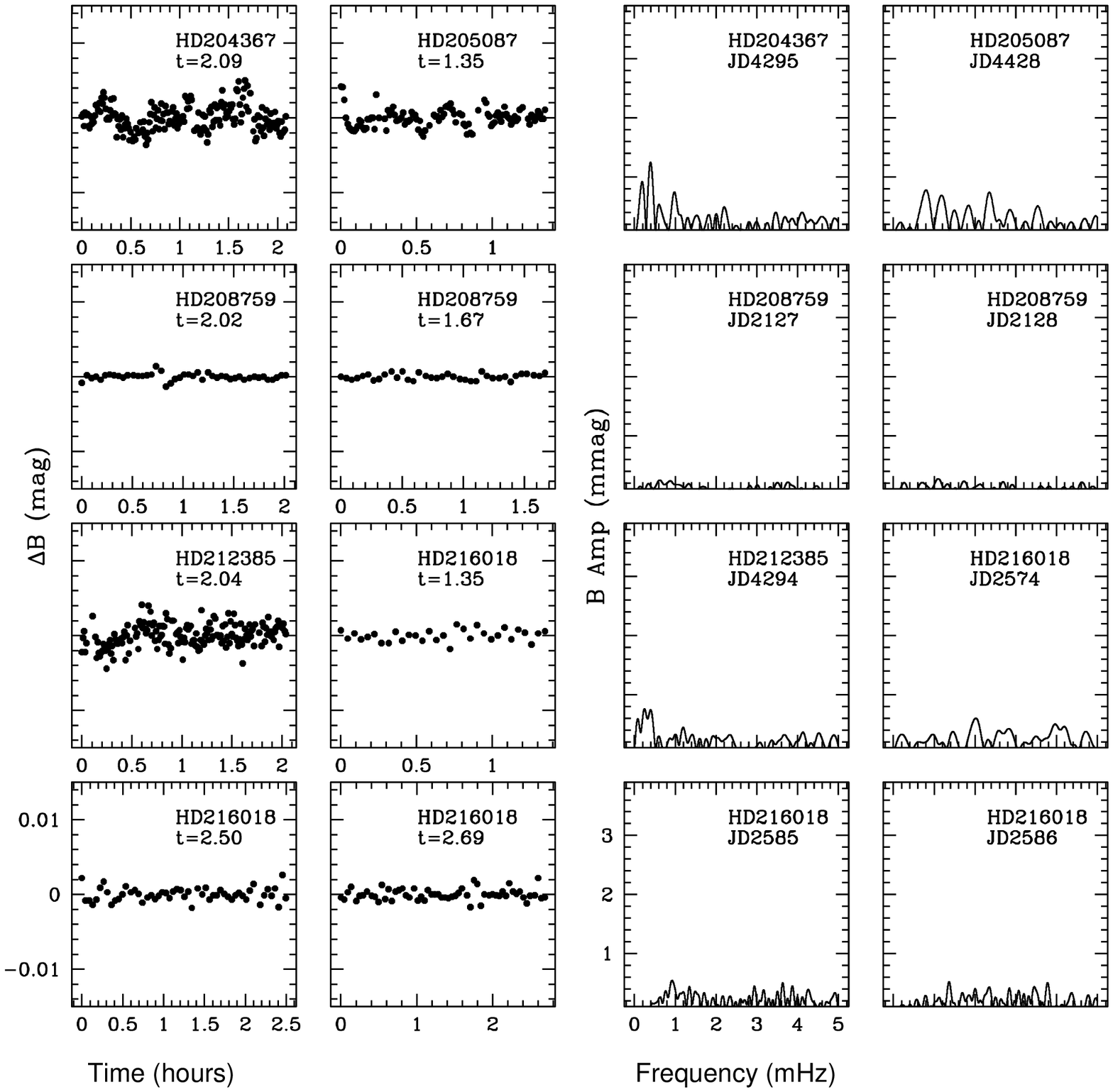}
    \caption{Continued.}
   \end{figure}
   
   \addtocounter{figure}{-1}
  \begin{figure}
   \centering
    \includegraphics[width=0.95\linewidth]{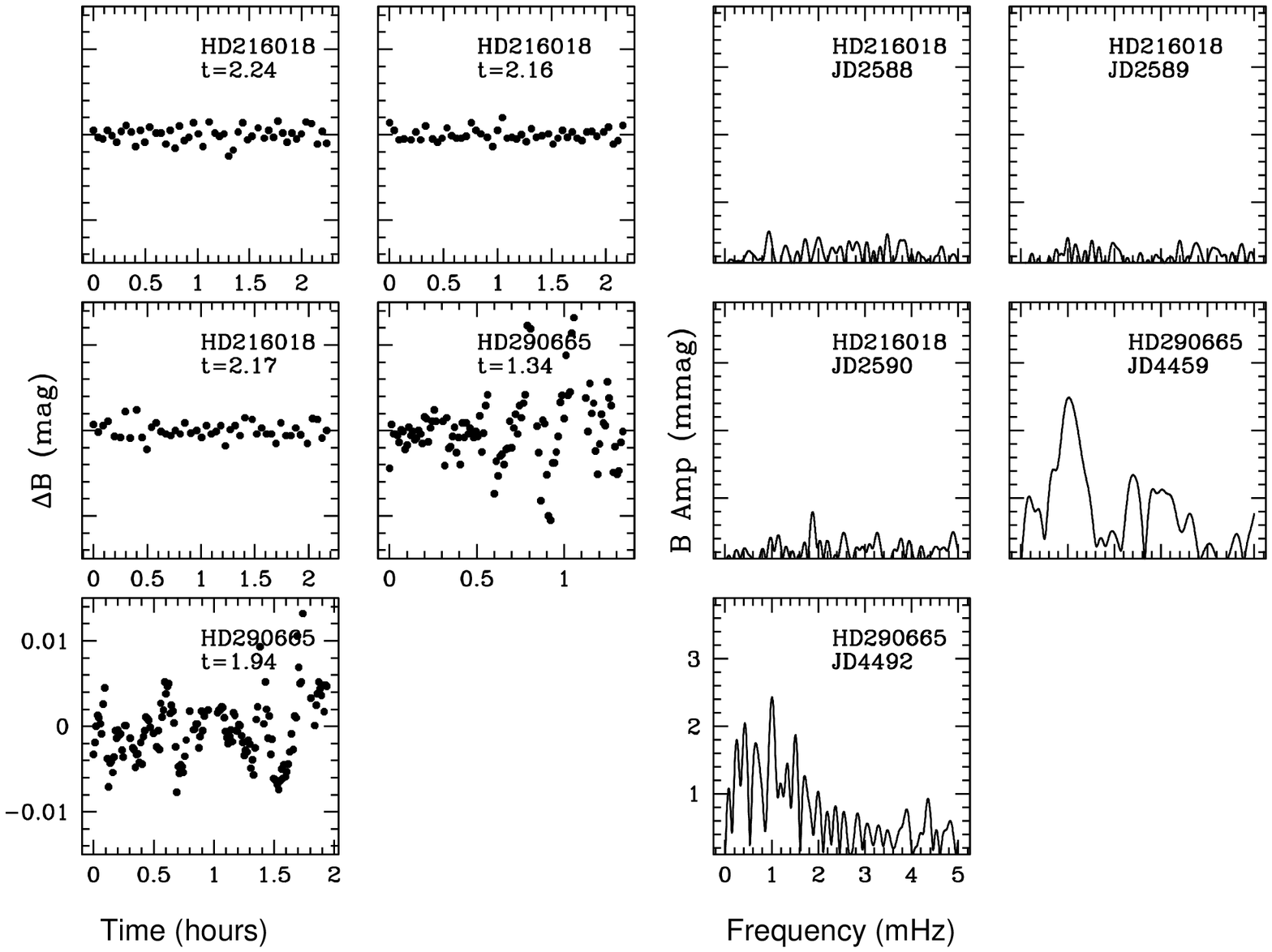}
    \caption{Continued.}
\end{figure}


\begin{thebibliography}{10}
\bibitem[1962]{abt62}Abt, H. A., \& Golson, J. C. 1962, ApJ, 136, 35
\bibitem[1973]{abt73}Abt, H. A., \& Snowden, M. S. 1973, ApJS, 25, 137
\bibitem[2012]{alen12}Alentiev, D., Kochukhov, O., Ryabchikova, T., et al. 2012, MNRAS, 421, L82
\bibitem[2001]{ashoka01}Ashoka, B. N., Kumar, Babu, V. C., et al. 2001, JA\&A, 22, 131
\bibitem[1992]{babe92}Babel, J. 1992, A\&A, 258, 449
\bibitem[2001]{balm01}Balmforth, N. J., Cunha, M. S., Dolez, N., et al. 2001, MNRAS, 323, 362
\bibitem[2003]{balo03}Balona, L. A., \& Laney, C. D. 2003, MNRAS, 344, 242
\bibitem[2011]{balo11}Balona, L. A., Cunha, M. S., Gruberbauer, M., et al. 2011, MNRAS, 413, 2651
\bibitem[2013]{balo13}Balona, L. A., Catanzaro, G., Crause, L., Cunha, M., et al. 2013, MNRAS, 432, 2808
\bibitem[2015]{balo15}Balona, L. A., Catanzaro, G., Abedigamba, O. P., et al. 2015, MNRAS, 448, 1378
\bibitem[1989]{belm89}Belmonte, J. A. 1989, Ph.D. Thesis, Universidad de La Laguna,  Tenerife
\bibitem[2003]{Bychkov03}Bychkov V. D., Bychkova L. V. \& Madej J. 2003, A\&A, 407, 631
\bibitem[2014]{cata14}Catanzaro, G., \&  Ripepi, V. 2014, MNRAS, 441, 1669
\bibitem[1993]{chris1993}Christensen-Dalsgaard J., 1993, Baglin, A., Weiss, W. W., eds, ASP Conf. Ser. Vol. 40, Proc. IAU Coll. 137, Inside the stars. ASP, San Francisco, 483
\bibitem[1999]{cox99}Cox, A. N. 1999, Allen's Astrophysical Quantities
\bibitem[1975]{craw75}Crawford, D. L. 1975, AJ, 80, 955
\bibitem[1979]{craw79}Crawford, D. L. 1979, AJ, 84, 1858
\bibitem[2001]{cunh01} Cunha, M. S., \& Gough, D. 2001, ESASP, 464, 419
\bibitem[1975]{deem75}Deeming, T. J. 1975, Ap\&SS, 36, 137
\bibitem[1998]{doro98}Dorokhova, T. N., \& Dorokhov, N. I. 1998, Contrib. Astron. Obser. Skalnate Pleso, 27, 338
\bibitem[2005a]{elki05a}Elkin, V. G., Riley, J. D., Cunha, M. S., Kurtz, D. W., Mathys, G. 2005a, MNRAS, 358, 665
\bibitem[2005b]{elki05b}Elkin, V. G., Kurtz, D. W., \& Mathys, G. 2005b, MNRAS, 364, 864
\bibitem[2010]{elki10}Elkin, V. G., Kurtz, D. W., Mathys, G., Freyhammer, L. M. 2010, MNRAS, 402, 1883
\bibitem[2011]{elki11}Elkin, V. G., Kurtz, D. W., Worters, H. L., et al. 2011, MNRAS, 411, 978
\bibitem[1996]{flow96}Flower, P. J. 1996, ApJ, 469, 355
\bibitem[1992]{gill92}Gilliland, R. L., \& Brown, T. M. 1992, PASP, 104, 582
\bibitem[1993]{gill93}Gilliland, R. L., Brown, T. M., Kjeldsen, H., et al 1993, AJ, 106, 2441
\bibitem[2001]{giri01}Girish, V., Seetha, S., Martinez, P., et al. 2001, A\&A, 380, 142
\bibitem[2008]{gonz08}Gonz\'alez, J. F., Hubrig, S., Kurtz, D. W., Elkin, V. G., Savanov, I. 2008, MNRAS, 384, 1140
\bibitem[1999]{hand99}Handler, G., \& Paunzen, E. 1999, A\&AS, 135, 57
\bibitem[2002]{hand02}Handler, G., Weiss, W. W., Paunzen, E., et al. 2002, MNRAS, 330, 153
\bibitem[2006]{hand06}Handler, G., Weiss, W. W., Shobbrook, R., et al. 2006, MNRAS, 366, 257
\bibitem[2004]{hatz04}Hatzes, A. P., \& Mkrtichian, D. E. 2004, MNRAS, 351, 2
\bibitem[2005]{hatz05}Hatzes, A. P., \& Mkrtichian, D. E. 2005, A\&A, 430, 279
\bibitem[1990]{hell90}Heller, C. H., \& Kramer, K. S. 1990, MNRAS, 244, 372
\bibitem[1992]{hild92}Hildebrandt, G. 1992, Astron. Nachr., 313, 233
\bibitem[2014a]{hold14a}Holdsworth, D. L., Smalley, B., Gillon, M., et al. 2014a, MNRAS, 439, 2078
\bibitem[2014b]{hold14b}Holdsworth D. L., Smalley, B., Kurtz, D. W., et al. 2014b, MNRAS, 443, 2049
\bibitem[2015]{hold15}Holdsworth, D. L. 2015, PhD Thesis, Keele University, UK
\bibitem[2015]{hou15}Hou, Wen, Luo, ALi, Yang, Haifeng, et al. 2015, MNRAS, 449, 1401
\bibitem[2012]{hubr12}Hubrig, S., Kurtz, D. W., Sch{\"o}ller M, et al. 2012, ASPC, 462, 318
\bibitem[2003]{josh03}Joshi, S., Girish, V., Sagar, R., et al. 2003, MNRAS, 344, 431
\bibitem[2006]{josh06}Joshi, S., Mary, D. L., Martinez, P., et al. 2006, A\&A, 455, 303
\bibitem[2009]{josh09}Joshi, S., Mary, D. L., Chakradhari, N. K., Tiwari, S. K., Billaud, C. 2009, A\&A, 507, 1763
\bibitem[2015]{josh15}Joshi, S., \& Joshi, Y. C. 2015, JA\&A, 36, 33
\bibitem[1998]{kana98}Kanaan, A., \& Hatzes, A. P. 1998, ApJ, 503, 848
\bibitem[1991]{kjel91}Kjeldsen, H., \& Frandsen, S. 1991, A\&AS, 87, 119
\bibitem[2010]{koch10}Koch, D. G., Borucki, W. J., Basri, G. et al. 2010, ApJ, 713, L79
\bibitem[2006]{koch06}Kochukhov, O. 2006, A\&A, 446, 1051
\bibitem[2001]{koch01}Kochukhov, O., \& Ryabchikova, T. 2001, A\&A, 374, 615
\bibitem[2008]{koch08}Kochukhov, O., Ryabchikova, T., Bagnulo, S., Lo Curto, G. 2008, A\&A, 479, L29
\bibitem[2009]{koch09}Kochukhov, O., Bagnulo, S., Lo Curto, G., Ryabchikova, T. 2009, A\&A, 493, L45
\bibitem[2013]{koch13}Kochukhov, O., Alentiev, D., Ryabchikova, T., et al. 2013, MNRAS, 431, 2808
\bibitem[2012]{korni12}Kornilov, V., Sarazin, M., Tokovinin, A., Travouillon, T., Voziakova, O. 2012, A\&A, 546, A41
\bibitem[1986]{krei86}Kreidl, T. J., \& Kurtz, D. W. 1986, MNRAS, 220, 313
\bibitem[1991]{krei91}Kreidl, T. J., Kurtz, D. W., Bus, S. J., et al. 1991, MNRAS, 250, 477
\bibitem[1978]{kurt78}Kurtz, D. W. 1978, IBVS, 1436, 1
\bibitem[1982]{kurt82}Kurtz, D. W. 1982, MNRAS, 200, 807
\bibitem[1985]{kurt85}Kurtz, D. W. 1985, MNRAS, 213, 773
\bibitem[1991]{kurt91}Kurtz, D. W. 1991, MNRAS, 249, 468
\bibitem[1987]{kurt87}Kurtz, D. W., \& Martinez, P. 1987, MNRAS, 226, 187
\bibitem[1994]{kurt94c}Kurtz, D. W., \& Martinez, P. 1994, IBVS, 4013, 1
\bibitem[1994a]{kurt94a}Kurtz, D. W., Sullivan, D. J., Martinez, P., Tripe, P. 1994a, MNRAS, 270, 674
\bibitem[1994b]{kurt94b}Kurtz, D. W., Martinez, P., \& Tripe, P. 1994b, MNRAS, 271, 421
\bibitem[1995]{kurt95}Kurtz, D. W., \& Martinez, P. 1995, IBVS, 4209, 1
\bibitem[1996]{kurt96}Kurtz, D. W., Martinez, P., Koen, C., Sullivan, D. J. 1996, MNRAS, 281, 883
\bibitem[1997a]{kurt97a}Kurtz, D. W., van Wyk, F., Roberts, G., et al. 1997a, MNRAS, 287, 69
\bibitem[1997b]{kurt97b}Kurtz, D. W., Martinez P., Tripe, P., Hanbury, A. G. 1997b, MNRAS, 289, 645
\bibitem[2003]{kurt03}Kurtz, D. W., Elkin, V. G., \& Mathys, G. 2003, MNRAS, 343, L5
\bibitem[2005a]{kurt05a}Kurtz, D. W., Elkin, V. G., \& Mathys, G. 2005a, MNRAS, 358, L6
\bibitem[2005b]{kurt05b}Kurtz, D. W., Elkin, V., Savanov, I., et al. 2005b, MNRAS, 358, 651
\bibitem[2006]{kurt06}Kurtz, D. W., Elkin, V. G., \& Mathys, G. 2006, MNRAS, 370, 1274
\bibitem[2011]{kurt11}Kurtz, D. W., Cunha, M. S., Saio, H.,  et al. 2011, MNRAS, 414, 2550
\bibitem[1990]{mart90a}Martinez, P. \& Kurtz, D. W. 1990,  MNRAS, 242, 636 
\bibitem[1990]{mart90b}Martinez, P., Kurtz, D. W., \& Heller, C. H. 1990, MNRAS, 246, 699
\bibitem[1991]{mart91}Martinez, P., Kurtz, D. W., \& Kauffmann, G. M. 1991, MNRAS, 250, 666
\bibitem[1993]{mart93}Martinez, P., Kurtz, D. W., \& Meitjes, P. J. 1993, MNRAS, 260, 9
\bibitem[1994a]{mart94a}Martinez, P., \& Kurtz, D. W. 1994a, MNRAS, 271, 118
\bibitem[1994b]{mart94b}Martinez, P., \& Kurtz, D. W. 1994b, MNRAS, 271, 129
\bibitem[1996]{mart96}Martinez, P., Weiss, W. W., Nelson, M. J., et al. 1996, MNRAS, 282, 243
\bibitem[1998]{mart98}Martinez, P., Meintjes, P., Ratcliff, S. J., Engelbrecht, C. 1998, A\&A, 334, 606
\bibitem[2001]{mart01}Martinez, P., Kurtz, D. W., Ashoka, B. N., et al. 2001, A\&A, 371, 1048
\bibitem[1987]{matt87}Matthews, J. M., Wehlau, W. H., \& Kurtz, D. W. 1987, ApJ, 313, 782
\bibitem[1988]{matt88}Matthews, J. M., Kreidl, T. J., \& Wehlau, W. H. 1988, PASP, 100, 255
\bibitem[2015]{medu15}Medupe, R., Kurtz, D. W., Elkin, V. G., Mguda, Z., Mathys, G. 2015, MNRAS, 446, 1347
\bibitem[1970]{micha70}Michaud, G. 1970, ApJ, 160, 641
\bibitem[1981]{micha81}Michaud, G., Charland, Y., \& Megessier, C. 1981, A\&A, 103, 244
\bibitem[2003]{mkrt03}Mkrtichian, D. E., Hatzes, A. P., \& Kanaan, A. 2003, MNRAS, 345, 781
\bibitem[2005a]{mkrt05a}Mkrtichian, D. E., \& Hatzes, A. P. 2005a, A\&A, 430, 263
\bibitem[2005b]{mkrt05b}Mkrtichian, D. E., \& Hatzes, A. P. 2005b, JApA, 26, 185 
\bibitem[1985]{moon85}Moon, T., \& Dworetsky, M. M. 1985, MNRAS, 217, 305
\bibitem[2015]{murp15}Murphy, S. J., Bedding, T. R., Niemczura, E., Kurtz, D. W., Smalley, B. 2015, MNRAS, 447, 3948
\bibitem[1993]{nels93}Nelson, M. J., \& Kreidl, T. J. 1993, AJ, 105, 1903
\bibitem[2015]{osb15}Osborn, J., F{\"o}hring, D., Dhillon, V. S., Wilson, R. W. 2015, MNRAS, 452, 1707
\bibitem[2012]{paun12}Paunzen, E., Netopil, M., Rode-Paunzen, M., et al. 2012, A\&A, 542A, 89
\bibitem[2015]{paun15}Paunzen, E., Netopil, M., Rode-Paunzen, M., et al. 2015, A\&A, 575A, 24
\bibitem[1974]{pres74}Preston, G. W. 1974, ARA\&A, 12, 257
\bibitem[2009]{rens09}Renson, P., \& Manfroid, J. 2009, A\&A, 498, 96
\bibitem[1991]{rens91}Renson P., Gerbaldi M., Catalano F. A., 1991, A\&AS, 89, 429
\bibitem[2000]{rich00}Richer, J., Michaud, G., \& Turcotte, S. 2000, ApJ, 529, 338
\bibitem[2004]{ryab04}Ryabchikova, T., Nesvacil, N., Weiss, W. W., Kochukhov, O., St{\"u}tz, Ch. 2004, A\&A, 423, 705
\bibitem[1992]{schn92}Schneider, H., Kreidl, T. J., \& Weiss, W. W. 1992, A\&A, 257, 130
\bibitem[1991]{schu91}Schutt, R. 1991, AJ, 101, 2177
\bibitem[1996]{Schwarzenberg96}Schwarzenberg-Cherny, A., 1996, ApJ, 460, 107
\bibitem[2011]{smal11}Smalley, B., Kurtz, D. W., Smith, A. M. S., et al. 2011, A\&A, 535, 3
\bibitem[2015]{smal15}Smalley, B., Niemczura, E., Murphy, S. J., et al. 2015, MNRAS, 452, 3334
\bibitem[2000]{turc00}Turcotte, S., Richer, J., Michaud, G., Christensen-Dalsgaard, J. 2000, A\&A, 360, 603
\bibitem[2007]{vanl07} van Leeuwen, F. 2007, \aap, 474, 653
\bibitem[1967]{youn67}Young, A. T. 1967, AJ, 72, 747
\bibitem[1974]{youn74}Young, A. T. 1974, ApJ, 189, 587
\end{thebibliography}
   \end{document}